\documentstyle[aps,preprint]{revtex}
\begin{document}
\preprint{IUCM94-021;\ CRPS-94-25}
\title{Collective modes of soliton-lattice states in double-quantum-well
systems}
\author{R. C\^ot\'e}
\address{D\'epartement de physique and Centre de
 Recherche en Physique du Solide,
Universit\'e de Sherbrooke, Sherbrooke, Qu\'ebec,  J1K 2R1, Canada.}
\author{L. Brey}
\address{Departamento Fisica Materia Condensada, Universidad Aut\'onoma,
Cantoblanco 28049 Madrid, Spain}
\author{H. Fertig}
\address{Department of Physics and Astronomy,University of
Kentucky,Lexington,Kentucky 40506-0055}
\author{A. H. MacDonald}
\address{Department of Physics, Indiana University,
 Bloomington, Indiana 47405}
\date{\today }
\maketitle

\begin{abstract}

In strong perpendicular magnetic fields double-quantum-well systems
can sometimes occur in unusual broken symmetry states which have
interwell phase coherence in the absence of interwell hopping.
When hopping is present in such systems
 and the magnetic field is tilted away from
the normal to the quantum well planes, a related
soliton-lattice state can occur which has kinks in the
dependence of the relative phase between electrons in opposite layers
on the coordinate perpendicular to the in-plane component of the
magnetic field.  In this article we evaluate the collective modes of this
soliton-lattice state in the generalized random-phase aproximation.
We find that, in addition to the Goldstone modes associated with the
broken translational symmetry of the soliton-lattice state, higher
energy collective modes occur which are closely related to the
Goldstone modes present in the spontaneously phase-coherent state.
We study the evolution of these collective
modes as a function of the strength of the in-plane magnetic field
and comment on the possibility of using the in-plane field
to generate a finite wave probe of the spontaneously phase-coherent
state.

\end{abstract}
\noindent
Pacs: 73.20.Dx,64.70.Rh,73.20.Mf

\section{Introduction}

Recently, there have been a number of experimental\cite{exp}
and theoretical\cite{theory}
investigations of the integer and fractional
quantum Hall effects in double-quantum-well systems (DQWS). The
additional degree of freedom due to the presence of the second well makes
the phase diagram of the two-dimensional electron gas in the DQWS very rich.
Technological progress has opened the possibility of producing DQWS's of very
high mobility where the spacing between the wells is small enough ($d\sim 100\
\AA $ ) to be comparable with the typical electron spacing within a layer.
In this regime a DQWS may
differ qualitatively from a pair of isolated single-layer systems.
In particular, theoretical work based on
several apparently different points of view
has established\cite{wenandzee,ezawa,collecmodes}
that for sufficiently small separations,
a novel broken symmetry can occur in DQWS's in a strong perpendicular
magnetic field
in which phase coherence exists between electrons in different layers even
when no
tunneling occurs.  This broken symmetry is favored by interlayer
electron-electron interactions because good interlayer correlations
are guaranteed by the Pauli exclusion principle when the phase
difference between the two layers is fixed.
(If the phase relationship is fixed, two electrons cannot share the
same planar coordinate even if they are in opposite layers.)
  The broken symmetry occurs at strong
perpendicular magnetic fields because Landau level degeneracy
allows the symmetry to be broken without a kinetic energy cost.
The broken symmetry is expected to exist in the ground state
for layer separation, $d$, smaller than a critical layer separation, $d_c$.
Moreover, for $d < d_c$ the system is expected\cite{wenandzee,dllong1}
to have a finite
 temperature phase transition of the Kosterlitz-Thouless type.

For realistic DQWS's with $d < d_c$ there is always a finite amplitude
for tunneling between the two layers of the DQWS.  Tunneling
favors a particular phase relationship between electrons in
opposite layers so that the symmetry discussed above is explicitly broken
by the tunneling term in the Hamiltonian and the Kosterlitz-Thouless phase
transition associated with its breaking does not occur.
However, recent experiments\cite{murphy}
have demonstrated that a second related quantum phase transition
occurs in DQWS's when the magnetic field is tilted away from
the normal to the layers of the DQWS.  When the component of the magnetic
field which is parallel to the electron layers is finite, electrons
must acquire an Aharonov-Bohm phase when they move around
paths which enclose flux from the parallel field.  In one convenient
gauge, the Aharonov-Bohm phase appears through phase
factors in the interlayer tunneling amplitudes which depend
on the planar coordinate which is perpendicular to the in-plane
component of the magnetic field.  The tunneling term in the Hamiltonian
then favors a phase relationship between electrons in opposite layers which
follows the phase of the tunneling amplitude.  When the phase
relationship is spatially dependent it becomes possible for electrons
in opposite layers to share the same planar coordinate so that the
interaction energy increases.

According to Yang {\it et al.}\cite{kunyang,dllong1,dllong2}
the phase transition seen by Murphy {\it et al.}\cite{murphy}
results from this competition between tunneling energy and
interaction energy in a parallel field.  In the simplest view,
the state on the small parallel field side of the transition is one where
the phase relationship between the layers follows the phase of the
tunneling amplitude while the state on the
large parallel field side of the transition is one
in which
the phase is independent of the spatial coordinate and the
symmetry-breaking tunneling term has no effect on the
ground-state energy of the DQWS.
A more accurate picture of this state
and of the phase transition
follows from the effective Hamiltonian derived by Yang {\it et al.}
in which the local interaction energy cost of spatial dependence in
the phase relationship between the layers is approximated by a
gradient term in the energy density.  Minimizing the energy in a
parallel field then gives rise to a sine-Gordon equation.
The state on the large parallel field side of
this transition is a soliton-lattice state\cite{kunyang,dllong2} which
has its translational symmetry broken by the introduction
of periodic kinks in the phase relationship between
electrons in the two wells of the DQWS as the parallel magnetic field is
increased.
The effective Hamiltonian derived by Yang {\it et al.}
may be mapped to thoroughly studied models (see for example
Refs.\cite{bak,gruner})
of commensurate-incommensurate phase transitions with the
parameter which tunes the transition proportional
to the parallel component of the magnetic field (see below).
Motivated by this mapping, we refer to the state on the
weak parallel field side of the transition
as the commensurate state and to the state
in the absence of tunneling
as the incommensurate state.

In this paper we use a generalized random-phase approximation (GRPA)
to study the response functions and collective modes of the
soliton-lattice state.  As discussed elsewhere\cite{dllong1} and in further
details
below, the GRPA is expected to be qualitatively correct for DQWS's at
total Landau level filling factor $\nu =1$.  It is at this filling
factor that the physical effects associated with spontaneous phase
coherence are most easily observed.  The use of a microscopic
theory frees us from the gradient approximation and, more importantly,
allows us
to study higher energy excitations which may not be accurately
rendered by the effective Hamiltonian.  In Section II, we detail
the microscopic Hamiltonian on which we base our calculations.
We appeal to the strong perpendicular magnetic field in limiting
the Hilbert space to the lowest Landau level in each layer and
assume that the electrons are completely spin polarized.
 GRPA response functions
are readily calculated from the Hartree-Fock approximation ground
state of the DQWS.  In Section III, we outline the
Hartree-Fock formalism which allows us to describe both
the simpler commensurate state and the more intricate
soliton-lattice state.  Numerical Hartree-Fock approximation results
for the ground state are compared with expectations based on the
gradient approximation effective Hamiltonian and with the theory of
commensurate-incommensurate phase transitions.  At strong
parallel magnetic fields corrections are quantitatively important.
We discuss results for the period of the soliton-lattice, the shape
of individual solitons, and the ground-state energy.
We show explicitly that the soliton-lattice state always has
lower energy than the incommensurate phase, although the energy
difference is extremely small at very strong
parallel magnetic fields.  We also calculate the critical strength
of parallel magnetic field at which the phase transition between
commensurate and soliton-lattice states occurs.
In Section IV we describe the GRPA calculations.  The response
functions of the soliton-lattice state are complicated by its
broken translational symmetry but we are able to derive
formal results which are convenient for numerical calculations by
taking advantage of simplifications which are due to the truncation
of the Hilbert space to the lowest Landau level.  The formalism is a
generalization to double-layers of one developed\cite{cotemacdo}
to describe the Wigner crystal state of single layer systems in
strong perpendicular magnetic fields and has been applied
previously\cite{cotebrey} to double-layer systems
in the case where there is no parallel field.  We find that the
lowest energy collective modes are Goldstone modes associated
with the broken translational symmetry of the soliton-lattice state.
Collective modes occur at higher energies which are closely related to
the Goldstone modes of the phase-coherent broken-symmetry state
in the absence of tunneling between the layers.  We comment
on the possibility of using a parallel magnetic field to
generate a finite wave vector probe of the coherent broken symmetry state.
Section V contains a brief summary of our results.

\section{Model Hamiltonian for DQWS's in a tilted magnetic field}

We consider a symmetric DQWS subjected to a tilted magnetic field.  We are
interested in studying the ground-state properties of this system as a
function of
both the parallel component of the magnetic field and the
separation $d$ (from center to center) between the wells at filling factor $%
\nu =1$.  We assume throughout this paper that the
perpendicular component of the magnetic field is kept constant.  The total
field is taken as ${\bf B}=B_{||\ }{\bf y}+B_{\perp }{\bf z}$ and, in the
Landau gauge, the vector potential ${\bf A}=(0,B_{_{\perp }\ }x,{\bf -}%
B_{||}x)${\bf . }We consider only the case of an unbiased DQWS:
the subband energies in the two layers are identical and,
in the ground state, the charge is equally distributed between the two layers.

Taking the perpendicular component of the magnetic field to be strong, we
restrict the Hilbert space to the first Landau level and assume the
electron gas to be fully spin polarized.  The many-body Hilbert space
then consists of occupation number states based on a set
of single-particle orbitals labeled by a layer index $i$ and a
guiding center index $X$:
\begin{equation}
\phi _{i,X}({\bf r})=\frac 1{\sqrt{%
L_y}}\frac 1{(\pi \ell ^2)^{1/4}}e^{-iXy/\ell ^2}e^{-(x-X)^2/2\ell ^2}\chi
_i(z),
\label{eq:sporb}
\end{equation}
where $\ell ^2=\hbar c/eB_{\perp }$ is the square of the magnetic length
for the {\it perpendicular} component of the magnetic field and $\chi
_i(z) $ with $i=R,L$ is the envelope wave function of the lowest-energy
state centered on the right or left well.  We can define
a separate magnetic length associated with the parallel component
of the magnetic field, $\ell_{||}^2 = \hbar c /e B_{|| }$.  In the absence
of interactions and tunneling between the wells these single particle
orbitals are eigenstates of the single-particle Hamiltonian.

For the sake of limiting the number of parameters characterizing our
DQWS, we make a narrow well approximation {\it i.e.} we assume that the
width, $b$, of the wells is small ($ b<< d$) and treat interlayer hopping in a
tight-binding\cite{hu} approximation.  The single-particle problem is then
characterized by the interlayer distance $d$ and a hopping integral $t$:
\begin{equation}
\label{atrois}H_0=\sum_{i,X}\epsilon \ c_{i,X}^{\dagger }c_{i,X}-\widetilde{t%
}\sum_X\left( e^{-iXd/\ell _{||}^2\ }c_{R,X}^{\dagger }c_{L,X}+h.c.\right).
\end{equation}
In this approximation, the parallel component of the magnetic field is
responsible for a reduction\cite{hu} in the magnitude of the
hopping integral
($\widetilde{t} =te^{-d^2\ell ^2/4\ell _{||}^4}\equiv te^{-Q^2\ell ^2/4}$)
and for the introduction of a guiding-center-dependent phase factor,
$e^{-iXd/\ell _{||}^2 } \equiv e^{-i Q X}$.
$\epsilon$ is the identical subband energy of
the quantum wells (including the lowest Landau-level energy
$\hbar\omega_c/2$, where $\omega_c=eB_\perp/mc$) and $Q$ is defined by
\begin{equation}
\label{aquatre}Q \equiv \frac d{\ell _{||}^2} \propto B_{||}.
\end{equation}
The parallel component of the magnetic field enters only in the
non-interacting part of the Hamiltonian but, as we will see, the
competition it produces with the interaction energy leads to interesting
many-body physics.

\section{Hartree-Fock ground states in a parallel field}

\subsection{Hamiltonian of the interacting 2D electron gas}

In order to evaluate the GRPA response functions we must first complete
self-consistent Hartree-Fock calculations for the ground state of
the DQWS.  The order parameters that describe the commensurate and
soliton-lattice states are defined below and we compute them using the
formalism
developed in Refs.\cite{cotebrey} and \cite{cotemacdo}.  Although this approach
is not the simplest approach one could use in the present context, it has the
advantage of being easy to implement numerically.  More importantly,
the results we derive here will lead to a remarkable simplification of the
GRPA response function calculations.

We start be defining the operator
\begin{equation}
\label{acinq}\rho _{ij}({\bf q})=\frac 1g\sum_Xe^{-iq_xX-iq_xq_y\ell ^2/2}\
c_{i,X}^{\dagger }c_{j,X+q_y\ell ^2},
\end{equation}
where $g=S/2\pi \ell ^2$ is the degeneracy of the Landau levels ($S$ is the
area of the system).   The single-particle Hartree-Fock Hamiltonian
can be expressed in terms of this operator.  Defining
${\bf Q}\equiv Q\widehat{x}$ we obtain
\begin{equation}
\label{asix}
\begin{array}{rcl}
H_{HF} & = & g\epsilon \rho (0)-g
\widetilde{t}\left( \rho _{RL}({\bf Q})+\rho _{LR}(-{\bf Q})\right) \\  & +
& g\left( \frac{e^2}{\epsilon _0\ell }\right) \sum_{i,j}\sum_{{\bf q}}\left[
H_{iijj}({\bf q})<\rho _{ii}(-{\bf q})>\rho _{jj}({\bf q})-X_{iijj}({\bf q}%
)<\rho _{ij}(-{\bf q})>\rho _{ji}({\bf q})\right] ,
\end{array}
\end{equation}
where
\begin{equation}
\label{asept}
\begin{array}{rll}
H_{iiii}({\bf q})=V_a ({\bf q}) & = & \left(
\frac 1{q\ell }\right) e^{-q^2\ell ^2/2}, \\ H_{iijj}({\bf q})=V_c({\bf q})
& = & \left(
\frac 1{q\ell }\right) e^{-q^2\ell ^2/2}e^{-qd}, \\ X_{iiii}({\bf q})=V_b(%
{\bf q}) & = & \int_0^\infty d(q^{\prime }\ell )J_0(qq^{\prime }\ell
^2)e^{-q^{\prime 2}\ell ^2/2}, \\
X_{iijj}({\bf q})=V_d({\bf q}) & = & \int_0^\infty d(q^{\prime }\ell
)J_0(qq^{\prime }\ell ^2)e^{-q^{\prime 2}\ell ^2/2}e^{-q^{\prime }d},
\end{array}
\end{equation}
are respectively the Hartree intrawell and interwell interactions and the Fock
intrawell and interwell interactions. These functions of wave vector depend
only
on the interlayer separation and are plotted in
Ref.\cite{cotebrey}. $J_0$ is a Bessel function of
the first kind and $\rho (0)=\rho
_{RR}(0)+\rho _{LL}(0)\;$is the  ``density'' operator.  In deriving these
results we have neglected processes in which electrons are scattered from
one well to the other as the result of their mutual interactions.  This
assumption is consistent with the tight-binding approximation.  In
realistic systems, small terms of this type will be present but they
do not add anything new to the physics.

These equations are very general and allow both positional and coherence
broken symmetries.
The set of quantities $<\rho _{ij}({\bf q})>$ describes all possible states
of the DQWS and we use them as order parameters: $<\rho _{RR}({\bf q})>$ and
$<\rho _{LL}({\bf q})>$ describe density fluctuations in each well while the
interwell coherence is reflected in a non-zero value of the interwell
order parameters $<\rho _{RL}({\bf q})>$ and $<\rho _{LR}({\bf q})>$ . The
states that we consider here are ones where the electron density is
uniform in each well, but the interwell order parameters are allowed to vary
in space along the $\hat x$ axis.  We do not allow variation in the
interwell order parameters along the $\widehat{y}$ direction.
In our notation, we restrict ourselves to solutions of the Hartree-Fock
equations where
\begin{equation}
\label{ahuit}
\begin{array}{rcl}
<\rho _{RR}(0)> & = & <\rho _{LL}(0)>=
\frac 12, \\ <\rho _{RL}(q_x,0)> & , & <\rho _{LR}(q_x,0)>\neq 0,
\end{array}
\end{equation}
and all other order parameters are zero.  It seems clear that the lowest-energy
solution of the Hartree-Fock equations satisfies these
restrictions for the situations of physical interest to us.  The
restriction that $<\rho _{RR}(0)>  =  <\rho _{LL}(0)>= 1/2$ is enforced
by the large Hartree energy cost of having unequal densities in the two
layers.  The only spatial dependence which we allow is that driven by the
hopping term in the Hamiltonian which depends on the $\hat x$ coordinate.
In the absence of a parallel field, we know\cite{cotebrey} that the
minimum-energy solution of the Hartree-Fock equations is spatially
uniform at small layer separations but breaks translational symmetry
by forming a Wigner crystal state at sufficiently large layer separations.
Once the Wigner crystal state is formed, interlayer coherence is quickly
lost.  For $\nu = 1$ the occurrence of a Wigner crystal state is believed to
be an artifact of the Hartree-Fock approximation.  In any event, our
present interest is in the regime of small layer separations where we do
not expect
spatial dependence other than that driven by the parallel field; here
we are confident that the lowest-energy Hartree-Fock state satisfies the
restrictions of Eq.(~\ref{ahuit}).  With these restrictions the
Hartree-Fock Hamiltonian simplifies:
\begin{equation}
\label{aneuf}
\begin{array}{rcl}
H_{HF} & = & g\left( \epsilon -
\frac 12\left( \frac{e^2}{\epsilon _0\ell }\right) V_b(0)\right) \rho (0)-g%
\widetilde{t}\left( \rho _{RL}({\bf Q})+\rho _{LR}(-{\bf Q})\right) \\  & -
& g\left( \frac{e^2}{\epsilon _0\ell }\right) \sum_{q_x}V_d(q_x)\left[ <\rho
_{RL}(-q_x)>\rho _{LR}(q_x)+<\rho _{LR}(-q_x)>\rho _{RL}(q_x)\right] ,
\end{array}
\end{equation}
where $V_b(0)=\sqrt{\frac \pi 2}$.  To arrive at Eq.(\ref{aneuf}), we have
added a neutralizing positive background; the energy $\epsilon$ appearing
in Eq.(\ref{aneuf}) now depends on the location of this background in three
dimensional space.  For constant perpendicular magnetic field,
the first term in the right-hand-side of Eq.(\ref{aneuf}) is a
constant which we absorb into the zero of energy in what follows.

\subsection{Equation of motion of the single-particle Green's function}

The Hartree-Fock Hamiltonian depends on order parameters
which must be determined self-consistently.  We now derive
explicit self-consistent equations by considering the equation of
motion of the single-particle Green's function
defined by $G_{ij}(X,X^{^{\prime
}},\tau )=-<Tc_{i,X}(\tau )c_{j,X^{^{\prime }}}^{\dagger }(0)>$ .
Within the restricted Hartree-Fock approximation discussed above, the
single-particle Green's function is diagonal in its guiding
center indices and we can work\cite{comment1}
with a one-component Fourier transform defined by
\begin{equation}
\label{aonze}G_{ij}(q_x,\tau )=\frac 1g\sum_XG_{ij}(X,X,\tau
)e^{-iq_xX}.
\end{equation}
The order parameters are related to this Green's function by $<\rho
_{ij}(q_x)>=G_{ji}(q_x,\tau =0^{-}).$   The equation of motion
of the Green's function is derived using $\hbar \frac
\partial {\partial \tau }(\cdots )=\left[ H-\mu N,\cdots \right] $
where $\mu $ is the chemical potential of the electron gas
and we set $\hbar =1$.  We find that
\begin{equation}
\label{adouze}
\begin{array}{rcl}
(i\omega _n+\mu )G_{RR}(q_x,\omega _n)+\widetilde{t}G_{LR}(q_x+Q,\omega _n)
& = & \delta _{q_{x,}0}-\sum_{q_x^{^{\prime }}}U^{*}(q_x^{^{\prime
}})G_{LR}(q_x+q_x^{^{\prime }},\omega _n) \\
(i\omega _n+\mu )G_{LR}(q_x,\omega _n)+\widetilde{t}G_{RR}(q_x-Q,\omega _n)
& = & -\sum_{q_x^{^{\prime }}}U(q_x^{^{\prime }})G_{RR}(q_x-q_x^{^{\prime
}},\omega _n),
\end{array}
\end{equation}
where the effective Hartree-Fock mean field is due to the
interlayer exchange energy and is given by
\begin{equation}
U(q_x)=\left( \frac{%
e^2}{\epsilon _0\ell }\right) V_d(q_x)<\rho _{RL}(q_x)>.
\label{aone}
\end{equation}

Our restricted Hartree-Fock equations do not couple different guiding centers.
Since we require that the charge density in each layer be uniform and
independent of position, it follows that for the Hartree Fock
eigenfunctions at guiding center $X$, the magnitudes of the
coefficients multiplying the basis functions localized in left and right
wells must be equal.  The only freedom of physical
relevance is the relative phase, $\theta (X)$, multiplying the basis functions.
In terms of this quantity,
\begin{equation}
\label{aqorze}<\rho _{RL}(q_x)>=\frac 1g\sum_X<\rho (X)>e^{-iq_xX}=\frac 1g%
\sum_X\ \left[ \frac 12e^{i\theta (X)}\right] \ e^{-iq_xX},
\end{equation}
where $<\rho (X)>=<c_{R,X}^{\dagger }c_{L,X}>$ is the interwell order
parameter in $X$ space.

At this point, it is more convenient\cite{kunyang,dllong2}
to define a new phase $\widetilde{%
\theta }(X)$ by
\begin{equation}
\label{aquinze}\widetilde{\theta }(X)=\theta (X)-QX.
\end{equation}
In the commensurate state, $\widetilde{\theta }(X) \equiv 0$,
while in the soliton-lattice state it is a periodic function of
$X$.  It is advantageous to make use of this periodicity
and redefine the interwell order parameter as
\begin{equation}
\label{aseize}<\widetilde{\rho }_{RL}(q_x)>\equiv <\rho _{RL}(q_x+Q)>=\frac 1%
{2g}\sum_Xe^{-iq_xX}e^{i\widetilde{\theta }(X)},
\end{equation}
which is the Fourier transform of $<\widetilde{\rho }(X)>=<\rho (X)>e^{-iQX}=%
\frac 12e^{i\widetilde{\theta }(X)}$ .

In our notation the commensurate-incommensurate
phase transition of Ref.\cite{kunyang} is described in the following way.
The {\it commensurate} phase (C phase) at weak parallel magnetic fields has
$\theta (X)=QX$; the phase of the wavefunction follows the phase of
the hopping matrix element just as it would for non-interacting
electrons.  For this state $\widetilde{\theta }(X)=0$ and
so only $<\widetilde{\rho }_{RL}(0)>=\frac 12$ is different from zero. The
commensurate phase has an energy per particle
\begin{equation}
\label{nou}E_C=-\widetilde{t}-\frac 14\left( \frac{e^2}{\epsilon _0\ell }%
\right) V_d(Q),
\end{equation}
which increases monotonically with magnetic field.
In the opposite limit of strong parallel magnetic fields, the phase does not
follow the imposed periodicity and we expect $\theta (X)=0$. Thus,
$\widetilde{%
\theta }(X)=-QX$ or, equivalently, only $<\widetilde{\rho }_{RL}(-{\bf Q)}>=%
\frac 12$ is non-zero. The energy per particle in the {\it incommensurate}
phase
(I phase) is
\begin{equation}
\label{aseizepp}E_I=-\frac 14\left( \frac{e^2}{\epsilon _0\ell }\right)
V_d(0),
\end{equation}
and is clearly independent of the parallel magnetic field.  There is a
critical parallel magnetic field $B_{||,(C\rightarrow I)\text{,}}$ at
which the energy of the commensurate state rises above the energy
of the incommensurate state.  As pointed out in Ref.\cite{kunyang},
however, before this critical magnetic field
is reached it becomes energetically favorable to introduce kinks (solitons)
where $ \widetilde{\theta }(X)$ changes by $2 \pi $; in the ground state
the solitons are arranged periodically.
We refer to this state as the {\it soliton-lattice} state (S phase).
For very strong parallel magnetic fields the solitons are smooth and
spaced by $2 \pi / Q $ so that $ \widetilde{\theta }(X) \approx -QX$
and the soliton-lattice state asymptotically approaches the
incommensurate state.  In Ref.\cite{kunyang} this phase transition
is described using a gradient approximation for the exchange
energy which allows the ground-state problem to be mapped to well-studied
models of commensurate-incommensurate phase transitions\cite{bak,gruner}.
In the Appendix, we briefly review the most relevant results from these
models and discuss the relationship between the gradient-approximation
theory for the ground state and the Hartree-Fock approximation.
At the critical parallel field for the transition between
commensurate and soliton-lattice states, $B_{||,(C\rightarrow S)}$,
the system admits a single soliton and
$\widetilde{\theta }(X)$ is a solution of the
sine-Gordon equation. (See the appendix for further details.)  As the
parallel magnetic field increases above this transition,
the solitons proliferate and it is necessary to take their
interactions into account.  This is best done numerically\cite{bak,gruner}
even in the gradient approximation.  Below we describe our numerical
Hartree-Fock calculations for the soliton-lattice ground state.

In the ground state of the soliton-lattice phase, we have the
periodicity $e^{i \widetilde{%
\theta }(X+L_S)}=e^{i\widetilde{\theta }(X)}$ where $L_S$, the period
of the soliton-lattice, is chosen to minimize the energy.
It follows that if we define a soliton wave vector by
\begin{equation}
\label{adsept}Q_S=\frac{2\pi }{L_S},
\end{equation}
the only non-zero order
parameters are $<\widetilde{\rho }_{RL}(nQ_S)>,\;n=0,\pm
1,\pm 2,\ldots $.  (We can choose the origin for the $\hat x$ coordinate
so that these order parameters are real.)
We can simplify the notation of Eqs.(\ref{adouze}) by redefining the interwell
single-particle Green's function as
\begin{equation}
\label{adhuit}\widetilde{G}_{LR}(q_x,\omega _n)\equiv G_{LR}(q_x+Q,\omega
_n).
\end{equation}
Defining $\ \widetilde{G}_{LR}(n,\omega _n)\equiv
\widetilde{G}_{LR}(nQ_S,\omega
_n),G_{RR}(n,\omega _n)\equiv G_{RR}(nQ_S,\omega _n),\ \widehat{u}(n)\equiv
\delta _{n,0}$ and the real matrix
\begin{equation}
\label{adhuitp}F(n,m)\equiv \widetilde{t}\delta _{n,m}+\left( \frac{e^2}{%
\epsilon _0\ell }\right) V_d((n-m)Q_S+Q)<\widetilde{\rho }_{RL}(n-m)>,
\end{equation}
we have, in an obvious matrix notation,
\begin{equation}
\label{adneuf}
\begin{array}{rcl}
(i\omega _n+\mu )G_{RR}(\omega _n)+F^T\widetilde{G}_{LR}(\omega _n) & = &
\widehat{u}, \\ (i\omega _n+\mu )\widetilde{G}_{LR}(\omega
_n)+FG_{RR}(\omega _n) & = & 0.
\end{array}
\end{equation}
These equations are readily decoupled:
\begin{equation}
\label{avun}
\begin{array}{rcl}
\left[ \left( i\omega _n+\mu \right) ^2I-B\right] G_{RR}(\omega _n) & = &
\left( i\omega _n
+\mu \right) \widehat{u}, \\ \left[ \left( i\omega
_n+\mu \right) ^2I-B\right] \widetilde{G}_{LR}(\omega _n) & = & -F\widehat{u}%
,
\end{array}
\end{equation}
where the real and symmetric matrix $B$ is defined by
\begin{equation}
\label{avdeux}B=F^TF=FF^T.
\end{equation}

$B$ can be diagonalized by an orthogonal transformation: $B=U\Omega U^T$
where $\Omega $ is the diagonal matrix of the eigenvalues of $B$ with
elements $\Omega _i.$ Solving for the order parameters, we find at finite
temperature
\begin{equation}
\label{avtrois}\left\langle \rho _{RR}(n)\right\rangle =\frac 12%
\sum_iU_{n,i}U_{i,0}^T\left[ f(E_{i,-})+f(E_{i,+})\right] ,
\end{equation}
and
\begin{equation}
\label{avquatre}
\begin{array}{rcl}
\left\langle \widetilde{\rho }_{RL}(n)\right\rangle & = & \frac 12\sum_{i,k}%
\frac{U_{n,i}U_{i,k}^TF(k,0)}{|E_i|}\left[
f(E_{i,-})-f(E_{i,+})\right] ,
\end{array}
\end{equation}
where $f(x)=\frac 1{e^{\beta (x-\mu )}+1}$ is the Fermi function and
\begin{equation}
\label{avcinq}E_{i,\pm }=\pm \sqrt{\Omega _i}.
\end{equation}
In order to avoid density fluctuations at $T=0K$, {\it i.e. }in order to
have $\left\langle \rho _{ii}(0)\right\rangle =\frac 12,\left\langle \rho
_{ii}(n\neq 0)\right\rangle =0\ (i=R,L)$ , we must have $\mu (T=0) =0$ so that
the Fermi factors equal one for the negative energy solutions and
zero for the positive energy solutions.
Then $\left\langle \rho
_{ii}(n)\right\rangle =\frac 12
\sum_jU_{n,j}U_{j,0}^{\dagger }=\frac 12\delta _{n,0}$
and, in matrix notation,
\begin{equation}
\label{avsix}\left\langle \widetilde{\rho }_{RL}\right\rangle =\frac 12%
U\Omega ^{-\frac 12}U^TF\widehat{{\bf u}},
\end{equation}
where $\Omega^{1/2}$ represents the diagonal matrix of
$\Omega_i^{1/2}$ values.
{}From this last equation, we can directly show that the interwell order
parameter obeys the sum rule
\begin{equation}
\label{sum}
\begin{array}{rcl}
\sum_n\left| \left\langle \widetilde{\rho }_{RL}(n)\right\rangle \right| ^2
& = & \left\langle
\widetilde{\rho }_{RL}\right\rangle ^T\left\langle \widetilde{\rho }%
_{RL}\right\rangle , \\  & = & \left(
\frac 12\widehat{{\bf u}}^TF^TU\Omega ^{-\frac 12}U^T\right) \left(
\frac 12U\Omega ^{-\frac 12}U^TF\widehat{{\bf u}}\right) , \\  & =
& \frac 14\widehat{{\bf u}}^TF^TB^{-1}F\widehat{{\bf u}}, \\  & = & \frac 14%
{}.
\end{array}
\end{equation}

Eqs.(\ref{avdeux}),(\ref{avcinq}),(\ref{avsix}) represent a set of
self-consistent, coupled equations that we must solve in an iterative way
until a stable solution is found for a given value of $Q_S,Q$ and $t$. The
calculation must then be repeated for different values of $Q_S$ until we
find the solution which minimizes the HF ground-state energy per particle
\begin{equation}
\label{avhuit}E=-2\widetilde{t}<\widetilde{\rho }_{RL}(0)>-\left( \frac{e^2}{%
\epsilon _0\ell }\right) \sum_nV_d(nQ_S+Q)\left\langle \widetilde{\rho }%
_{RL}(n)\right\rangle ^2.
\end{equation}
Given the order parameters, we can easily find $\widetilde{\theta
}(X)$ from the equation
\begin{equation}
\label{xspace}<\widetilde{\rho }_{RL}(X)>=\frac 12e^{i\widetilde{\theta }%
(X)}=\sum_ne^{inQ_SX}<\widetilde{\rho }_{RL}(n)>.
\end{equation}
The Hartree-Fock eigenvalues
$E_{\pm}\left( X\right) $ (see appendix) are related to the
order parameters by
\begin{equation}
\label{band}E_{\pm }(X)=\pm \left| \widetilde{t}+\sum_nV_d(nQ_S+Q)<%
\widetilde{\rho }_{RL}(n)>e^{inQ_SX}\right| .
\end{equation}

Commensurate and incommensurate states also represent extrema of
the Hartree-Fock energy functional and therefore are self-consistent
solutions of the Hartree-Fock equations.
For commensurate and incommensurate states respectively we find that
\begin{equation}
\label{band1}E_C^{\pm }(X)=\pm \left( \widetilde{t}+\frac 12V_d(Q)\right) ,
\end{equation}
and
\begin{equation}
\label{band2}E_I^{\pm }(X)=\pm \left| \widetilde{t}+\frac 12%
V_d(0)e^{-iQX}\right| .
\end{equation}

\subsection{Numerical results for the Hartree-Fock Approximation Ground
State}

In Fig. 1, we plot the energies of the C, I and S phases for (a) $t/\left(
e^2/\ell \right) =0.01,d/\ell =1.0$ and (b) $t/\left( e^2/\ell \right)
=0.005,$\ $d/\ell =1.877$. The dotted line represents the gradient
approximation for the commensurate state energy
\begin{equation}
\label{grad}E_{C,(grad)}\simeq -\widetilde{t}-\frac 14\left( \frac{e^2}{%
\epsilon _0\ell }\right) V_d(0)+\pi \ell ^2\rho _SQ^2.
\end{equation}
The stiffness $\rho _S$ is easily obtained from $V_d(Q)$ using Eq.(\ref{ros}%
).  We find for (a) $\pi\rho _S=0.0195\ \left( e^2/\ell \right) $ and (b)
$\pi\rho
_S=0.00751\left( e^2/\ell \right) $.  The gradient approximation is very
close to the HF energy for $E_C$.  Analytic results
from the gradient approximation for the
the values of $Q$ at which the commensurate and
incommensurate states are equal in energy ($Q_{C\rightarrow I}$)
and at which the soliton state first becomes
stable ($Q_{C\rightarrow S}$) are also in good agreement with the the HF
values.
Numerically, we find that for (a) $Q_{C\rightarrow S}/Q_{C\rightarrow
I}=0.91$, and
for (b) $Q_{C\rightarrow S}/Q_{C\rightarrow I}=0.95$ while in the gradient
approximation, this ratio is $Q_{C\rightarrow S}/Q_{C\rightarrow I}=2\sqrt{%
2}/\pi =0.90$.

We also see in Fig.~1 that the energy of the soliton phase
is always lower than the energy of the incommensurate phase so that
this phase is never thermodynamically stable in the Hartree-Fock
approximation. Of course, as $Q$ becomes very large, the order parameters in
the soliton phase become asymptotically close to those of the $I$ phase and
the energies of the two phases becomes nearly identical. One should also notice
that the difference in energy between the soliton and incommensurate or
commensurate phase is of the order of 1\% at most!

The behavior of $\widetilde{\theta }(X)$ over one period of the soliton
lattice is shown in Fig.~2.
As expected, the width of the soliton is very large right at the
transition and then decreases as the parallel magnetic field increases.
In the large $Q$ limit, $\widetilde{\theta }(X)\rightarrow -QX+\pi $, which is
the incommensurate solution (the constant $\pi $ here comes from the fact
that we have made an overall phase convention which forces the
soliton solution to vary from
$2\pi $ to $0$ over the interval $(-L_S/2,L_S/2)$).

The dotted line in (a) and (b) represents the sine-Gordon
solution $\widetilde{\theta }(X)=4\tan ^{-1}\left[ e^{-X/\xi }\right] $,
where $\xi=\sqrt{2\pi\rho_S\ell^2/{\tilde t}}$ (see Eq.(\ref{xsi})).
Fig.~3, is a plot of the soliton wave vector $ Q_S \equiv 2 \pi / L_S$
as a function of $Q\ell $.  As
the parallel magnetic field increases, the soliton wave vector
asymptotically approaches $Q$ as required to approach the
incommensurate state solution.

Fig.~4 shows the shape of the upper band $E_{+}(X)$ over one period $L_S$
for two extreme values of $Q\ell $.  The dispersion of the soliton band
structure seen in (a) weakens as the parallel magnetic field is
increased (see (b)). In the Hartree-Fock approximation the charge gap
at $\nu = 1$, the minimum energy to make a particle-hole pair, is the
sum of the maximum and the minimum values of these bands.  (Remember
that the Hartree-Fock eigenvalues at $\nu = 1$ occur in opposite
energy pairs.) It is the charge gap which is measured in the
transport\cite{murphy} experiments.
We see in Fig.~4 that, within the Hartree-Fock approximation,
the effect of interactions on the charge gap
is enormous.  The hopping energy is here
of order $0.01\left( e^2/\ell \right) $ while the band energy is one order
of magnitude higher!  However, we find that the Hartree-Fock charge gap is
only weakly
dependent on the parallel component of the magnetic field, in complete
disagreement with experiment.  This behavior is actually expected.
The lowest-energy charged excitations are
believed\cite{kunyang,dllong1,dllong2,kymacd}
to be spin-textures and not the simple single-particle excitations
whose energies are obtained from the Hartree-Fock calculations.
The spin textures in double-layer systems are
analogous to the skyrmion spin-textures which are
known to be the lowest energy charged excitations\cite{skyr,fertig} of a
single-layer system which is not spin-polarized.

Fig.~5, shows the commensurate energy for various values of the bare hopping
$t$. If we approximate the critical wave vector for the commensurate-soliton
transition by that for the $C\rightarrow I$ transition (this is not a bad
approximation as we saw above), we find the curve given in Fig.~6. Notice
that because of the renormalization of $t$ to $\widetilde{t}$, there is
always a critical value of $Q$ at which this transition occurs.
The angle that the magnetic field
makes with the DQWS at the transition is given by
$\tan (\theta_c )=Q\ell /\left( d/\ell \right).$
In Fig.~6, we plot the critical angle for the transition at $d/\ell =1$.
It is clear that the critical angle has a
square-root dependence on $t$ at small $t$. The same result was\cite{kunyang}
obtained in the gradient approximation.  Experimentally the critical tilt
angle is identified with a feature in the dependence of the charge
gap on tilt angle and it appears\cite{murphy} that the tilt angle
has a linear dependence on $t$.
Moreover, for given $t$ the critical angle appears to be overestimated
by both the Hartree-Fock approximation and by the gradient approximation
when the stiffness coefficient, $\rho_S$, (see Eq.(\ref{ros})) is
calculated in the Hartree-Fock
approximation.\cite{kunyang}
For example, the critical angle that would be
given for the parameters $d=1.877,t=0.005$ (see Fig.1 (b) above) which are
similar to those for the sample
studied in Ref.\cite{murphy}, is approximately of $%
21^{\circ }$ . This value is about $2.6$ times greater than that of the
experiment ($8^{\circ }$).  As discussed elsewhere\cite{dllong2} we
believe that this quantitative discrepancy is due to quantum
fluctuations not included in the Hartree-Fock approximation.
For larger values of $t$ these fluctuations are suppressed
and we expect, in agreement with experiment, that critical tilt
angles will agree more closely with the Hartree-Fock approximation.

\section{Response functions of the commensurate and soliton phases in the
GRPA}

We now consider the calculation of the response functions of the DQWS in the
presence of the tilted magnetic field.  The total response is characterized
in general by sixteen coupled susceptibilities:
\begin{equation}
\label{un}\chi _{ijkl}({\bf q,q}^{\prime },\tau )=-g<T\widehat{\rho }_{ij}(%
{\bf q,\tau )}\widehat{\rho }_{kl}(-{\bf q}^{\prime }{\bf ,}0{\bf )>,}
\end{equation}
where $\widehat{\rho}_{ij}=\rho -<\rho_{ij}>$.
It turns out (as in the ground-state calculation), that the equations of
 motion for the susceptibilities have a
simpler form if we work with the shifted operators $\widetilde{\rho }_{RL,}%
\widetilde{\rho }_{LR}$ instead of the unshifted ones.
We define the $ 4 \times 4 $ matrix
\begin{equation}
\label{deux}
\begin{array}{lcc}
\Gamma  ({\bf q,q}^{\prime },\tau ) \equiv &  &  \\
&  &  \\
\left(
\begin{array}{cccc}
\chi _{RRRR}({\bf q,q}^{\prime },\tau ) & \chi _{RRLR}({\bf q,q}^{\prime }%
{\bf +Q},\tau ) & \chi _{RRRL}({\bf q,q}^{\prime }-{\bf Q},\tau ) & \chi
_{RRLL}(
{\bf q,q}^{\prime },\tau ) \\ \chi _{RLRR}({\bf q+Q,q}^{\prime },\tau ) &
\chi _{RLLR}({\bf q+Q,q}^{\prime }{\bf +Q},\tau ) & \chi _{RLRL}({\bf q+Q,q}%
^{\prime }-{\bf Q},\tau ) & \chi _{RLLL}(
{\bf q+Q,q}^{\prime },\tau ) \\ \chi _{LRRR}({\bf q-Q,q}^{\prime },\tau ) &
\chi _{LRLR}({\bf q-Q,q}^{\prime }{\bf +Q},\tau ) & \chi _{LRRL}({\bf q-Q,q}%
^{\prime }-{\bf Q},\tau ) & \chi _{LRLL}(
{\bf q-Q,q}^{\prime },\tau ) \\ \chi _{LLRR}({\bf q,q}^{\prime },\tau ) &
\chi _{LLLR}({\bf q,q}^{\prime }+{\bf Q},\tau ) & \chi _{LLRL}({\bf q,q}%
^{\prime }-{\bf Q},\tau ) & \chi _{LLLL}({\bf q,q}^{\prime },\tau )
\end{array}
\right) , &  &
\end{array}
\end{equation}
where again ${\bf Q=}Q\widehat{{\bf x}}$.

The generalized random-phase approximation (GRPA) for the 2DEG in a strong
magnetic field with broken translational symmetry is
discussed in detail in Ref.\cite{cotemacdo}. The application of this
formalism to the present problem presents no major difficulty although the
algebra is knotty.  Below we briefly sketch the derivation of
the expressions we use for the numerical evaluation of the GRPA response
functions.

We first define the functions

\begin{equation}
\label{quatre}
\begin{array}{rcl}
f(q_x) & = & \widetilde{t}\delta _{q_x,0}+\left( \frac{e^2}{\varepsilon
_0\ell }\right) V_d(q_x+Q)<\widetilde{\rho }_{RL}(q_x)>, \\ F({\bf q-q}%
^{\prime }) & = & f(q_x-q_x^{^{\prime }})e^{-i(q_x-q_x^{^{\prime
}}+Q)q_y\ell ^2/2}\ \delta _{q_{y,}q_y^{^{\prime }}}, \\
F^T({\bf q-q}^{\prime }) & = & F(
{\bf q}^{\prime }{\bf -q}), \\ K({\bf q-q}^{\prime }) & = & <
\widetilde{\rho }_{RL}(q_x-q_x^{^{\prime }})>e^{-i(q_x-q_x^{^{\prime
}}+Q)q_y\ell ^2/2}\ \delta _{q_{y,}q_y^{^{\prime }}}, \\ K^T({\bf q-q}%
^{\prime }) & = & K({\bf q}^{\prime }{\bf -q}),
\end{array}
\end{equation}
and the $4\times 4$ matrices

\begin{equation}
\label{cinq}P({\bf q,q}^{\prime })=\left[
\begin{array}{cccc}
0 & -F^T({\bf q-q}^{\prime }) & F({\bf q-q}^{\prime }) & 0 \\
-F^{*}({\bf q-q}^{\prime }) & 0 & 0 & F(
{\bf q-q}^{\prime }) \\ F^{\dagger }({\bf q-q}^{\prime }) & 0 & 0 & -F^T(
{\bf q-q}^{\prime }) \\ 0 & F^{\dagger }({\bf q-q}^{\prime }) & -F^{*}({\bf %
q-q}^{\prime }) & 0
\end{array}
\right] ,
\end{equation}

\begin{equation}
\label{six}S({\bf q,q}^{\prime })=\left[
\begin{array}{cccc}
0 & -K^T({\bf q-q}^{\prime }) & K({\bf q-q}^{\prime }) & 0 \\
-K^{*}({\bf q-q}^{\prime }) & 0 & 0 & K(
{\bf q-q}^{\prime }) \\ K^{\dagger }({\bf q-q}^{\prime }) & 0 & 0 & -K^T(
{\bf q-q}^{\prime }) \\ 0 & K^{\dagger }({\bf q-q}^{\prime }) & -K^{*}({\bf %
q-q}^{\prime }) & 0
\end{array}
\right] ,
\end{equation}

\begin{equation}
\label{sept}X({\bf q,q}^{\prime })=\left[
\begin{array}{cccc}
V_b({\bf q})\delta _{{\bf q,q}^{\prime }} & 0 & 0 & 0 \\
0 & V_d({\bf q+Q)}\delta _{{\bf q,q}^{\prime }} & 0 & 0 \\
0 & 0 & V_d({\bf q-Q)}\delta _{{\bf q,q}^{\prime }} & 0 \\
0 & 0 & 0 & V_b({\bf q})\delta _{{\bf q,q}^{\prime }}
\end{array}
\right] ,
\end{equation}
and

\begin{equation}
\label{huit}H({\bf q,q}^{\prime })=\left[
\begin{array}{cccc}
V_a({\bf q})\delta _{{\bf q,q}^{\prime }} & 0 & 0 & V_c(
{\bf q})\delta _{{\bf q,q}^{\prime }} \\ 0 & 0 & 0 & 0 \\
0 & 0 & 0 & 0 \\
V_c({\bf q})\delta _{{\bf q,q}^{\prime }} & 0 & 0 & V_a({\bf q})\delta _{%
{\bf q,q}^{\prime }}
\end{array}
\right] .
\end{equation}
Using $\hbar \frac \partial {\partial \tau }(\cdots )=\left[ H-\mu N,\cdots
\right] $ where $H$ is the Hartree-Fock Hamiltonian (Eq.(\ref{aneuf})), we
find that the bare susceptibility $\Gamma^0$ (corresponding to a one-loop
approximation with the propagators evaluated in the HFA) satisfies the
matrix equation

\begin{equation}
\label{neuf}\sum_{{\bf q}^{\prime \prime }}\left[ I(\omega +i\delta )\
\delta _{{\bf q,q}^{\prime \prime }}-P({\bf q,q}^{\prime \prime })\right]
\Gamma^0({\bf q}^{\prime \prime },{\bf q}^{\prime },\omega )=S({\bf q,q}%
^{\prime }).
\end{equation}
Here $I$ is a unit matrix.
We note that this bare susceptibility is independent of the value of $q_y$
since the propagators do not involve $q_y$. This is no longer the case when
we add bubbles and ladder diagrams.The summation of the ladder diagrams is
given by the equation ($\widehat{\Gamma }$ is the irreducible polarization)

\begin{equation}
\label{dix}\widehat{\Gamma }({\bf q,q}^{\prime },\omega )=\Gamma ^0({\bf q,q}%
^{\prime },\omega )-\sum_{{\bf q}^{\prime \prime },{\bf q}^{\prime \prime
\prime }}\Gamma ^0({\bf q,q}^{\prime \prime },\omega )X({\bf q}^{\prime \prime
},{\bf q}^{\prime \prime \prime })\widehat{\Gamma }({\bf q}^{\prime \prime
\prime },{\bf q}^{\prime },\omega ),
\end{equation}
so that the full GRPA\ susceptibility is a sum of bubbles diagrams
containing the irreducible polarization and is given by

\begin{equation}
\label{onze}\Gamma ({\bf q,q}^{\prime },\omega )=\widehat{\Gamma }({\bf q,q}%
^{\prime },\omega )+\sum_{{\bf q}^{\prime \prime },{\bf q}^{\prime \prime
\prime }}\widehat{\Gamma }({\bf q,q}^{\prime \prime },\omega )H({\bf q}%
^{\prime \prime },{\bf q}^{\prime \prime \prime })\Gamma ({\bf q}^{\prime
\prime \prime },{\bf q}^{\prime },\omega ).
\end{equation}
The $q_y$ dependence enters only through the interaction $H$ and $X$ and so
it is obvious that $\Gamma ({\bf q,q}^{\prime },\omega )=\Gamma (q_x{\bf ,}%
q_x^{\prime };q_y,\omega )\delta _{q_y,q_y{}^{\prime }}$.

Because of the periodicity of the system, we must have $\Gamma ({\bf q,q}%
^{\prime },\omega )=\Gamma ({\bf k+}n{\bf Q,k+}m{\bf Q},\omega )$, where $%
n,m=0,\pm 1,\pm 2,\ldots $ and ${\bf k}$ is a vector in the first Brillouin
zone of the modulated structure {\it i.e.} $k_x\in \left[ \frac{-\pi }{L_s},%
\frac \pi {L_s}\right] $ and $k_y$ is unrestricted (the Brillouin zone is
effectively one-dimensional). This implies that all elements in the above
matrices are themselves matrices with respect to the indices $n,m$. Eqs. (%
\ref{neuf})-(\ref{onze}) can then be written in a purely matrix form and the
structure of the GRPA becomes much more transparent. One must however be
very careful not to interchange the order of these matrices since they do
not commute! Combining these three equations, we arrive at
\begin{equation}
\label{douze}\left[ I(\omega +i\delta )\ -R\right] \Gamma =S,
\end{equation}
where
\begin{equation}
\label{treize}R=P-S(X-H),
\end{equation}
or, more explicitly
\begin{equation}
\label{qorze}R=\left[
\begin{array}{cccc}
0 & -F^T+K^TX_{22} & F-KX_{33} & 0 \\
-F^{*}+K^{*}X_{11}-K^{*}H_{11}+KH_{14} & 0 & 0 & -R_{21}^{*} \\
F^{\dagger }-K^{\dagger }X_{11}+K^{\dagger }H_{11}-K^TH_{14} & 0 & 0 &
-R_{31}^{*} \\
0 & -R_{12}^{*} & -R_{13}^{*} & 0
\end{array}
\right] .
\end{equation}
Looking at the form of the matrix $R$, it is quite clear that Eq.(\ref{douze}%
) can be further simplified. After some algebra, we find that
\begin{equation}
\label{quinze}\left[ I(\omega +i\delta )^2\ -\Lambda \right] \Gamma =\Xi ,
\end{equation}
where
\begin{equation}
\label{seize}\Lambda =\left[
\begin{array}{cccc}
\Lambda _{11} & 0 & 0 & \Lambda _{14} \\
0 & \Lambda _{22} & \Lambda _{23} & 0 \\
0 & \Lambda _{32} & \Lambda _{33} & 0 \\
\Lambda _{14}^{*} & 0 & 0 & \Lambda _{11}^{*}
\end{array}
\right] ,
\end{equation}
with
\begin{equation}
\label{dsept}
\begin{array}{rcl}
\Lambda _{11} & = & R_{12}R_{21}+R_{13}R_{31,} \\
\Lambda _{22} & = & R_{21}R_{12}+R_{21}^{*}R_{12}^{*}, \\
\Lambda _{33} & = & R_{31}R_{13}+R_{31}^{*}R_{13}^{*}, \\
\Lambda _{14} & = & -R_{12}R_{21}^{*}-R_{13}R_{31}^{*}, \\
\Lambda _{23} & = & R_{21}R_{13}+R_{21}^{*}R_{13}^{*}, \\
\Lambda _{32} & = & R_{31}R_{12}+R_{31}^{*}R_{12}^{*},
\end{array}
\end{equation}
and
\begin{equation}
\label{dhuit}\Xi =\left[
\begin{array}{cccc}
R_{12}K^{*}-R_{13}K^{\dagger } & -(\omega +i\delta )K^T & (\omega +i\delta
)K & -R_{12}K+R_{13}K^T \\
-(\omega +i\delta )K^{*} & R_{21}K^T+R_{21}^{*}K^{\dagger } &
-R_{21}K-R_{21}^{*}K^{*} & (\omega +i\delta )K \\
(\omega +i\delta )K^{\dagger } & R_{31}K^T+R_{31}^{*}K^{\dagger } &
-R_{31}K-R_{31}^{*}K^{*} & -(\omega +i\delta )K^T \\
-R_{12}^{*}K^{*}+R_{13}^{*}K^{\dagger } & (\omega +i\delta )K^{\dagger } &
-(\omega +i\delta )K^{*} & R_{12}^{*}K-R_{13}^{*}K^T
\end{array}
\right] .
\end{equation}

In order to make connection with the pseudospin language used in
Ref.\cite{kunyang}, we define the density and pseudospin operators by
(remember that
real spins are assumed fully polarized)

\begin{equation}
\label{transfo}
\begin{array}{ccc}
n & \equiv & \rho _{RR}+\rho _{LL}, \\
S_Z & \equiv & \frac 12\left( \rho _{RR}-\rho _{LL}\right) .
\end{array}
\end{equation}
For example, we can now construct the density response function as well as
the longitudinal and transverse pseudospin response functions
\begin{equation}
\label{vdeux}
\begin{array}{rcl}
\chi _{nn} & = & \chi _{RRRR}+\chi _{RRLL}+\chi _{LLRR}+\chi _{LLLL}, \\
\chi _{S_ZS_Z} & = & \frac 14\left( \chi _{RRRR}-\chi _{RRLL}-\chi
_{LLRR}+\chi _{LLLL}\right) , \\ \chi _{+\ -} & = & \chi _{RLLR}.
\end{array}
\end{equation}
For the functions of Eq.(\ref{vdeux}), we need to solve for
\begin{equation}
\label{dneuf}\chi _L=\left[
\begin{array}{cc}
\chi _{RRRR} & \chi _{RRLL} \\
\chi _{LLRR} & \chi _{LLLL}
\end{array}
\right] ,
\end{equation}
and
\begin{equation}
\label{vingt}\chi _T=\left[
\begin{array}{cc}
\chi _{RLLR} & \chi _{RLRL} \\
\chi _{LRLR} & \chi _{LRRL}
\end{array}
\right] .
\end{equation}
With the definitions
\begin{equation}
\label{vtrois}R_A=\left[
\begin{array}{cc}
R_{12} & R_{13} \\
-R_{12}^{*} & -R_{13}^{*}
\end{array}
\right] ,
\end{equation}
\begin{equation}
\label{vquatre}R_B=\left[
\begin{array}{cc}
R_{21} & -R_{21}^{*} \\
R_{31} & -R_{31}^{*}
\end{array}
\right] ,
\end{equation}
and
\begin{equation}
\label{vcinq}K_A=\left[
\begin{array}{cc}
K^{*} & -K \\
-K^{\dagger } & K^T
\end{array}
\right] ,
\end{equation}
we arrive at the form of the
equations which we have solved numerically:
\begin{equation}
\label{vsept}
\begin{array}{ccc}
\left[ I(\omega +i\delta )^2-R_AR_B\right] \chi _L & = & R_AK_A \\
\left[ I(\omega +i\delta )^2-R_BR_A\right] \chi _T & = & R_BK_A^{\dagger }.
\end{array}
\end{equation}
Notice that $R_{12\text{ }}$is not simply
related to $R_{21}\ $so that there is no simple relationship between $R_A$
and $R_B$. These two matrices do not commute in the general case. We stress
once again that $R_A$ and $R_B$ are not simple $2\times 2$ matrices since
each of their elements is itself an infinite matrix. Inverting Eq. (\ref
{vsept}), we easily arrive at
\begin{equation}
\label{vseptp}
\begin{array}{ccc}
\chi _L & = & R_A\left[ I(\omega +i\delta )^2-R_BR_A\right] ^{-1}K_A, \\
\chi _T & = & \left[ I(\omega +i\delta )^2-R_BR_A\right]
^{-1}R_BK_A^{\dagger }.
\end{array}
\end{equation}
Where now the order of the matrices $R_A$ and $R_{B\text{ }}$are the same in
both equations and the matrix $R_BR_A$ is real. This shows clearly that all
response functions do share the same poles although each pole will in
general have different residues for different response functions.
The collective modes
associated with the longitudinal response functions are thus the same as
those associated with the transverse response functions. We can of course
solve, in the same way, for all other response functions in Eq.(\ref{deux}).

If we diagonalize the matrix $R_BR_A$ by
\begin{equation}
\label{vsppp}\left( R_BR_A\right) U=U\Omega ,
\end{equation}
where $\Omega $ is a diagonal matrix of the eigenvalues and $U$ the matrix
of eigenvectors, we can write
\begin{equation}
\label{vsix}(\chi _T)_{ij}=\sum_{l,m,k}\frac{%
U_{il}(U^{-1})_{lm}(R_B)_{mk}(K_A^{\dagger })_{kj}}{(\omega +i\delta
)^2-\Omega _l}.
\end{equation}

When $q_y=0$, all matrices are real and the above equations simplify
considerably. In this case, we can easily show that
\begin{equation}
\label{tun}
\begin{array}{c}
\chi _{nn}=0, \\
\chi _{S_zS_z}=\ \chi _{RRRR}.
\end{array}
\end{equation}
There is thus no density response to potentials which are
independent of the coordinate along the parallel magnetic field.

The numerical procedure to compute the response functions and find the
collective excitations is very simple. Once the order parameters are known
from the HFA described in the previous section, we can compute the matrices $%
R_B,R_A,$ and $K_A.$ We then diagonalize the real matrix $R_BR_A$ to find
the eigenvectors $U$ and eigenvalues $\Omega .$ We can then compute the
susceptibility from Eq.(\ref{vsix}) or, by following the pole with large
weight as the momentum ${\bf k}${\bf \ }varies in the Brillouin
zone, find the dispersion relation of the collective modes.
Before describing our numerical results for the soliton-lattice state,
it is instructive to look at the response functions in the
commensurate phase where the GRPA equations can be solved exactly.

\subsection{Response functions in the commensurate phase}

In the commensurate phase, the only non-zero order parameter is $<\widetilde{%
\rho }_{RL}(0)>=1/2$. It follows then that

$$
\begin{array}{rcl}
f(q_x) & = & t_R\delta _{q_x,0}, \\
F({\bf q,q}^{\prime }) & = & t_Re^{-i\varphi }\ \delta _{
{\bf q},{\bf q}^{\prime }}, \\ F^T({\bf q,q}^{\prime }) & = & F(
{\bf q},{\bf q}^{\prime }), \\ K({\bf q,q}^{\prime }) & = & \frac 12%
e^{-i\varphi }\ \delta _{{\bf q},{\bf q}^{\prime }}, \\ K^T({\bf q,q}%
^{\prime }) & = & K({\bf q,q}^{\prime }),
\end{array}
$$
where we have defined

\begin{equation}
\label{tr}t_R\equiv \widetilde{t}+\frac 12V_d(Q),
\end{equation}
and

\begin{equation}
\label{phi}\varphi \equiv \frac{Qq_y\ell ^2}2.
\end{equation}
It is very simple to solve for $\Gamma ({\bf q,q}^{\prime },\omega )$ given in
Eq.(\ref{deux}). First, we use the pseudospin transformation of Eq.(\ref
{transfo}) to define a new susceptibility matrix by%
$$
\begin{array}{lcc}
\Gamma_p({\bf q,q}^{\prime },\omega )\equiv &  &  \\
&  &  \\
\left(
\begin{array}{cccc}
\chi _{nn}({\bf q,q}^{\prime },\omega ) & \chi _{n-}({\bf q,q}^{\prime }{\bf %
+Q},\omega ) & \chi _{n+}({\bf q,q}^{\prime }-{\bf Q},\omega ) & \chi
_{nS_z}(
{\bf q,q}^{\prime },\omega ) \\ \chi _{+n}({\bf q+Q,q}^{\prime },\omega ) &
\chi _{+-}({\bf q+Q,q}^{\prime }{\bf +Q},\omega ) & \chi _{++}({\bf q+Q,q}%
^{\prime }-{\bf Q},\omega ) & \chi _{+S_z}(
{\bf q+Q,q}^{\prime },\omega ) \\ \chi _{-n}({\bf q-Q,q}^{\prime },\omega )
& \chi _{--}({\bf q-Q,q}^{\prime }{\bf +Q},\omega ) & \chi _{-+}({\bf q-Q,q}%
^{\prime }-{\bf Q},\omega ) & \chi _{-S_z}(
{\bf q-Q,q}^{\prime },\omega ) \\ \chi _{S_zn}({\bf q,q}^{\prime },\omega )
& \chi _{S_z-}({\bf q,q}^{\prime }+{\bf Q},\omega ) & \chi _{S_z+}({\bf q,q}%
^{\prime }-{\bf Q},\omega ) & \chi _{S_zS_z}({\bf q,q}^{\prime },\omega )
\end{array}
\right) . &  &
\end{array}
$$
After some simple algebra, we find
\begin{equation}
\label{chicommen}\Gamma_p({\bf q,q}^{\prime },\omega )=\left( \frac{\delta _{%
{\bf q},{\bf q}^{\prime }}}{(\omega +i\delta )^2-\omega _0^2({\bf q})}%
\right) \left(
\begin{array}{cccc}
-2b\left( 1-\cos \left( 2\varphi \right) \right) & i\omega \sin \left(
\varphi \right) & -i\omega \sin \left( \varphi \right) & ib\sin \left(
2\varphi \right) \\
-i\omega \sin \left( \varphi \right) & -a & a & -{\omega\over 2} \cos
\left( \varphi
\right) \\
i\omega \sin \left( \varphi \right) & a & -a & {\omega\over 2}
 \cos \left( \varphi
\right)  \\
-ib\sin \left( 2\varphi \right) & -{\omega\over 2}
 \cos \left( \varphi \right) &
{\omega\over 2} \cos \left( \varphi \right) & -{b\over 2}\left( 1+\cos
\left( 2\varphi \right)
\right)
\end{array}
\right) ,
\end{equation}
where we have defined
\begin{equation}
\label{omegao}
\begin{array}{rcl}
a & = & t_R+
\frac 12\left( V_a({\bf q})-V_b({\bf q})-V_c({\bf q})\cos \left( 2\varphi
\right) \right) , \\ b & = & t_R-
\frac 14\left( V_d({\bf q}+{\bf Q})+V_d({\bf q}-{\bf Q})\right) , \\ \omega
_0^2({\bf q}) & = & 4ab.
\end{array}
\end{equation}

{ }From Eq.(\ref{chicommen}), we see that the modulation of the pseudospin
structure ({\it i.e.}, of $\widetilde{\theta }(X)$ ) due to the parallel
magnetic field induces a coupling between the charge and spin response
functions. In fact, as discussed in Ref.\cite{dllong1}, this coupling is due
to the magnetic field and technically arises
because we have restricted the Hilbert space to the
first Landau level only. In the restricted Hilbert space, pseudospin and
charge operators no longer commute and it follows that modulations of the
pseudospin texture are in general
accompanied by modulations of the charge density. The
presence of the parallel magnetic field is crucial, however, for this
coupling to show up. In fact, the density response function is non-zero only
in the presence of a parallel magnetic field. Without this field, the ground
state consists of a fully filled Landau level of symmetric states.
Since a scalar field coupling to the density does not change the
pseudospin state, there is
no way to excite an electron from the symmetric to the antisymmetric state.
When the parallel field is added, however, symmetric and antisymmetric
states are mixed and density wave excitations becomes possible. By density
wave excitation, we mean here a collective excitation {\it i.e.} a pole of the
two-particle Green's function.

The coupling between pseudospin and density modes is interesting
because the collective excitations of the system must now be considered as
mixed pseudospin and density excitations. The coupling should  also be
important
experimentally since external electromagnetic fields will tend to
couple only to the total density of the two layers; in the presence of
a parallel component of the magnetic field the density-density response
function has a pole at the frequency of the pseudospin collective mode.
In the soliton-lattice phase to which we turn next and,
which has gapless collective modes, this coupling is also present.
A similar situation is encountered
in spiral magnets where the non-collinearity of the spin structure gives
rise to a spin-charge coupling\cite{tremblay}.

\subsection{Numerical results for the collective excitations}

In the commensurate phase, the various response functions
 $\chi $ are very simple
since they have only one real pole (no broadening). The pole of the bare
response functions $\chi ^0$ corresponds to the excitation of an electron (a
particle-hole excitation) from the $E_{-}$ to $E_{+}$ band (Eq.(\ref{avcinq})%
). When vertex corrections are introduced in the GRPA, this pole is
renormalized into a dispersive collective excitation (a spin wave in the
pseudospin language) and appears as a pole of $\chi $ .
These vertex corrections are important. To give an example,
 the gap calculated from
$\omega
_0(q=0)$ (Eq.(\ref{omegao})), at $d/\ell=1.0$ and $t/\left(
e^2/\epsilon_0\ell \right)=0.01$,
is reduced by a factor six by the vertex
corrections.
Fig.~7  shows the dispersion
relation of this spin wave, $\omega _0({\bf q})$, for different values of
the parallel magnetic field along the direction $q_y=0$ (the dispersion
relation is not isotropic since the parallel magnetic field is set along the
$y$ axis). A soft mode develops as the parallel magnetic field is increased,
but the value of $Q$ at which the collective mode
of the commensurate state first goes soft is larger than the
value of $Q$ at which the commensurate to soliton-lattice phase
transition occurs.  The commensurate to soliton lattice
phase transition is first order.
No corresponding partial mode softening appears as a
precursor of the commensurate to soliton lattice
phase transition along the direction $q_x=0$.

The response functions of the soliton-lattice state are more interesting.
Fig.~8 shows the imaginary part of the transverse response function $\chi
_{+-}({\bf k}+{\bf Q},{\bf k}+{\bf Q},\omega)$ (a) in the
 HFA and (b) in the GRPA.   The renormalization of the
particle-hole excitation is more dramatic than in the commensurate phase.
In addition to a Goldstone
mode which appears at low energies, a number of additional
sharp peaks suggestive of further collective modes
appear in the response functions.  For wave vectors in the $\hat x$
direction and parallel fields
close to that of the commensurate to soliton lattice phase transition,
the Goldstone modes are best thought of
as compressional waves of the
soliton lattice.  For wave vectors in the $\hat y$ direction the Goldstone
modes
in this regime are best thought of as oscillations in the position of a
particular
phase slip as a function of the $\hat y$ coordinate.
This Goldstone mode is expected because of the translational invariance
of the underlying Hamiltonian which is broken in the soliton-lattice
state.  The dispersion relation of the Goldstone modes is highly anisotropic
as shown in Fig.~9  where it is
computed for different values of $Q$ along the two perpendicular
directions $q_x=0$ and $q_y=0$.  Close to the phase transition between
commensurate and soliton-lattice states the solitons are very widely
spaced and we should expect that the velocity of the Goldstone
mode in the $q_y=0$ direction should be very small.
In this direction (the direction perpendicular to $B_{||}$),
the Goldstone mode is periodic with the periodicity of the Brillouin zone%
{\it \ i.e.} $2\pi /L_S$.

As we mentioned above, {\it in the absence of tunneling}, the ground state
of the system is independent of parallel field,
has spontaneous inter-layer phase coherence, and has
Goldstone modes corresponding to this broken symmetry.
The incommensurate state is the Hartree-Fock approximation for the
broken symmetry state.  For non-zero values of the tunneling amplitude
in a parallel field,
either the commensurate or the soliton-lattice state has a lower energy
and the phase transition between these states occurs at
a critical parallel field.  For very
strong magnetic fields, however, the soliton-lattice state asymptotically
approaches the incommensurate state.  The Goldstone modes of the
soliton-lattice state, which are most naturally associated with
broken translational symmetry, must crossover to the Goldstone
modes of the spontaneous-phase-coherence state.  These are easily
obtained by solving
GRPA equations at $t = 0$. (The calculations are very similar to those
for the commensurate state).
We find immediately that the response functions have a single pole at a
frequency given by
\begin{equation}
\label{omegai}
\omega_{sc}(q)=\sqrt{[V_a(q)-V_b(q)-V_c(q)+V_d(0)][V_d(0)-V_d(q)]}.
\end{equation}
In Fig.~10 we compare the
collective modes of the soliton-lattice
state in the strong-field regime (where $Q_S\approx Q$)
with the collective modes of the spontaneously phase-coherent
state in the absence of tunneling.
As expected the collective modes are nearly identical.
For the soliton-lattice state many collective modes appear at
each wavevector; for  $(k_x,k_y)$ the energies
are close to $\omega_{sc}(|(k_x+nQ,k_y)|)$
where $n$ is an integer. (The higher energy collective modes in Fig.~8
correspond to $n \ne 0$.) The broken translational symmetry of
the soliton lattice state results in a folding of the collective
modes of the $t=0$ ground state.  At the parallel field of Fig.~10
mode-mode coupling effects due to the interlayer hopping
are weak; as the parallel field weakens and
the phase transition is approached, the coupling becomes stronger
and eventually the modes become more similar to those of the incommensurate
state, except at wavelengths approaching the lengthening soliton-lattice
period.

At strong parallel fields, the effect of the tunneling on
the spontaneously-coherent ground state is very much like the effect of a
grating with wavevector $Q$.
We are presently investigating the possibility\cite{avenir} of
using tunneling in a parallel field as a tunable grating
which allows light to couple to finite wave vector modes of the spontaneously
phase-coherent state in much the same way that artificially
generated metallic gratings are used to couple to finite wave vector
plasmons of two-dimensional electron systems.\cite{grating}

Finally, we have verified that the
pseudospin-charge coupling that we discussed in the commensurate phase
also appears in the soliton phase. However, the pseudospin-charge
coupling is very small. Typically, the density response function
is more than one-thousand times smaller than the pseudospin response functions
$\chi_{ZZ}$ or $\chi_{+-}$. The weight of the Goldstone mode
in the density response function (as in all other response functions)
vanishes as the wave vector of the excitation vanishes.
Further studies will be needed to understand how this
coupling affects the ground-state conductivity of the soliton lattice.

\section{Conclusion}

We have studied the commensurate-incommensurate
phase transition which appears in a DQWS in a strong magnetic
field when the field is tilted away from the normal to the planes
of the two-dimensional electron gas in each well.
Working in the Hartree-Fock approximation, we have calculated
the energy of the soliton-lattice state for different values
of the tunneling parameter and parallel magnetic field and found that the
soliton-lattice state is the ground-state of the system for
all angles greater than the critical tilt angle. The difference in
energy between the soliton-lattice state and the incommensurate state
is however very small. Close to
the commensurate-soliton lattice transition, we find that the
gradient approximation described in Ref.\cite{kunyang} is a good
approximation for the soliton-lattice energy and for the modulation
of the interlayer coherence phase. This approximation soon breaks down,
however,  as the
 parallel magnetic
field is increased.

Using the GRPA, we have also studied the collective modes of the soliton
lattice.
As expected from the
broken translational symmetry of the soliton lattice, there is
a Goldstone mode branch which represents the phonons of the soliton
lattice. Because of the one-dimensional character of the
Brillouin zone, the Goldstone mode dispersion relation
is periodic only
in the direction perpendicular to the parallel magnetic field.
Beside the Goldstone mode,
there are also higher-energy collective excitations in the response
functions calculated in the GRPA. In the limit of strong parallel
magnetic fields, we were able to relate these higher-energy excitations to the
collective modes of the spontaneously-coherent state in the absence
of tunneling at wavevectors displaced by multiples of
$Q \hat x$.  In this regime, the effect of tunneling in a parallel field
resembles that of a finite-wave-vector grating which mixes the
different modes of the spontaneously coherent state.

Finally, we have also shown that, except in the direction perpendicular
to the parallel magnetic field, the collective modes in the
system are in fact mixed pseudospin-charge excitations.  Since all
response functions are coupled in the soliton lattice, it follows that
the density response function has poles at exactly the frequency of
these mixed pseudospin-charge excitations (including the Goldstone
mode). More work is needed, however, to understand how this coupling
could affect the conductivity of the soliton-lattice state.

\section{Acknowledgement}

This research was supported in part by NATO Collaborative Research Grant No.
930684.  RC acknowledges support from the Natural Sciences
and Engineering Research Council of Canada
(NSERC) and from the Fonds pour la formation de chercheurs et l'aide \`a la
recherche from the Government of Qu\'ebec (FACR).
HF acknowledges support from NSF grant No. DMR 92-02255,
from the Sloan Foundation, and from the Research Corporation.
LB acknowledges support from CICyT of Spain under
contract No. MAT 91-0201.  AHM acknowledges support from NSF grant
No. DMR 94-16906.  The authors acknowledge helpful discussions with
Kun Yang and Steve Girvin.

\vfill\eject

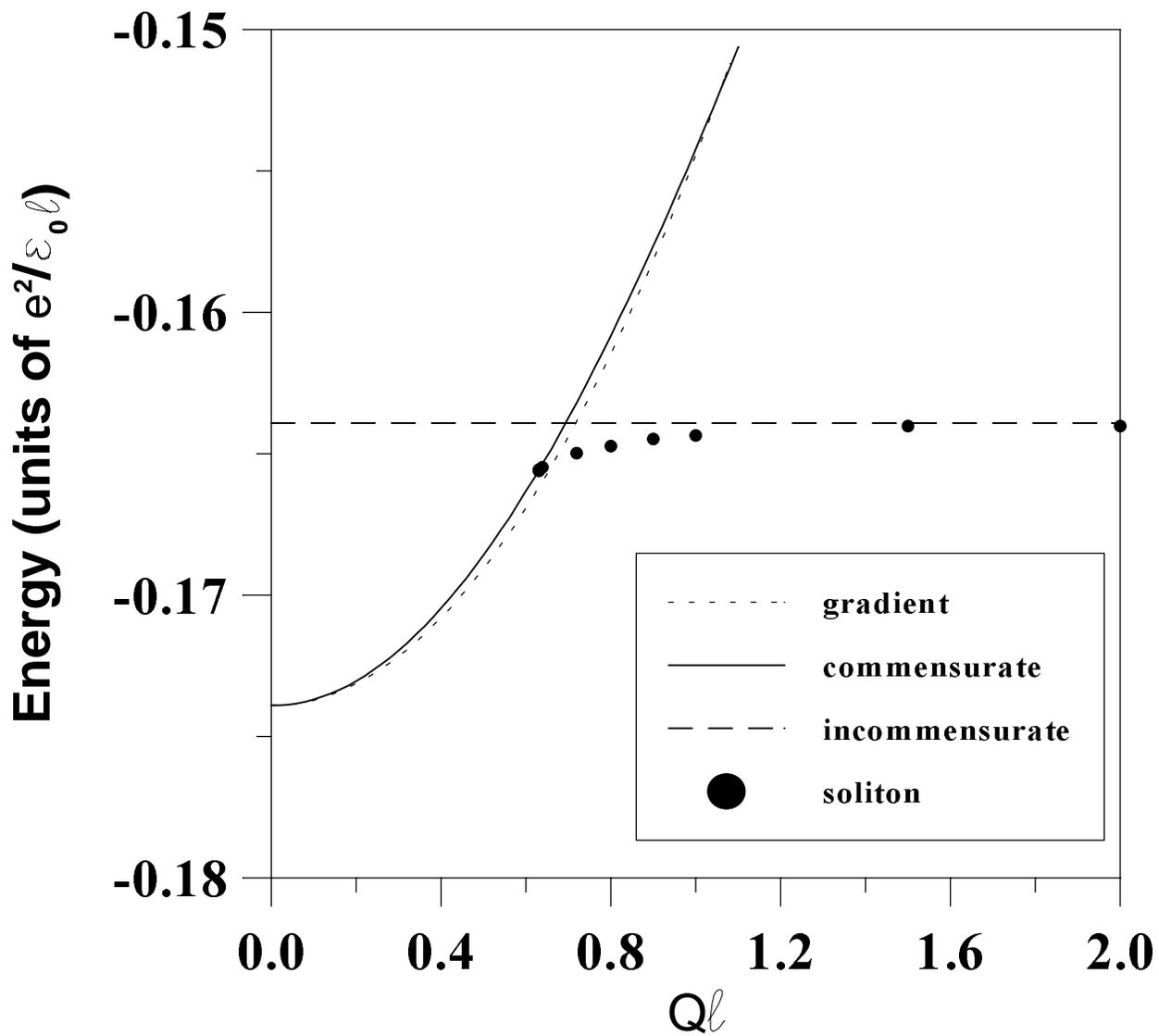
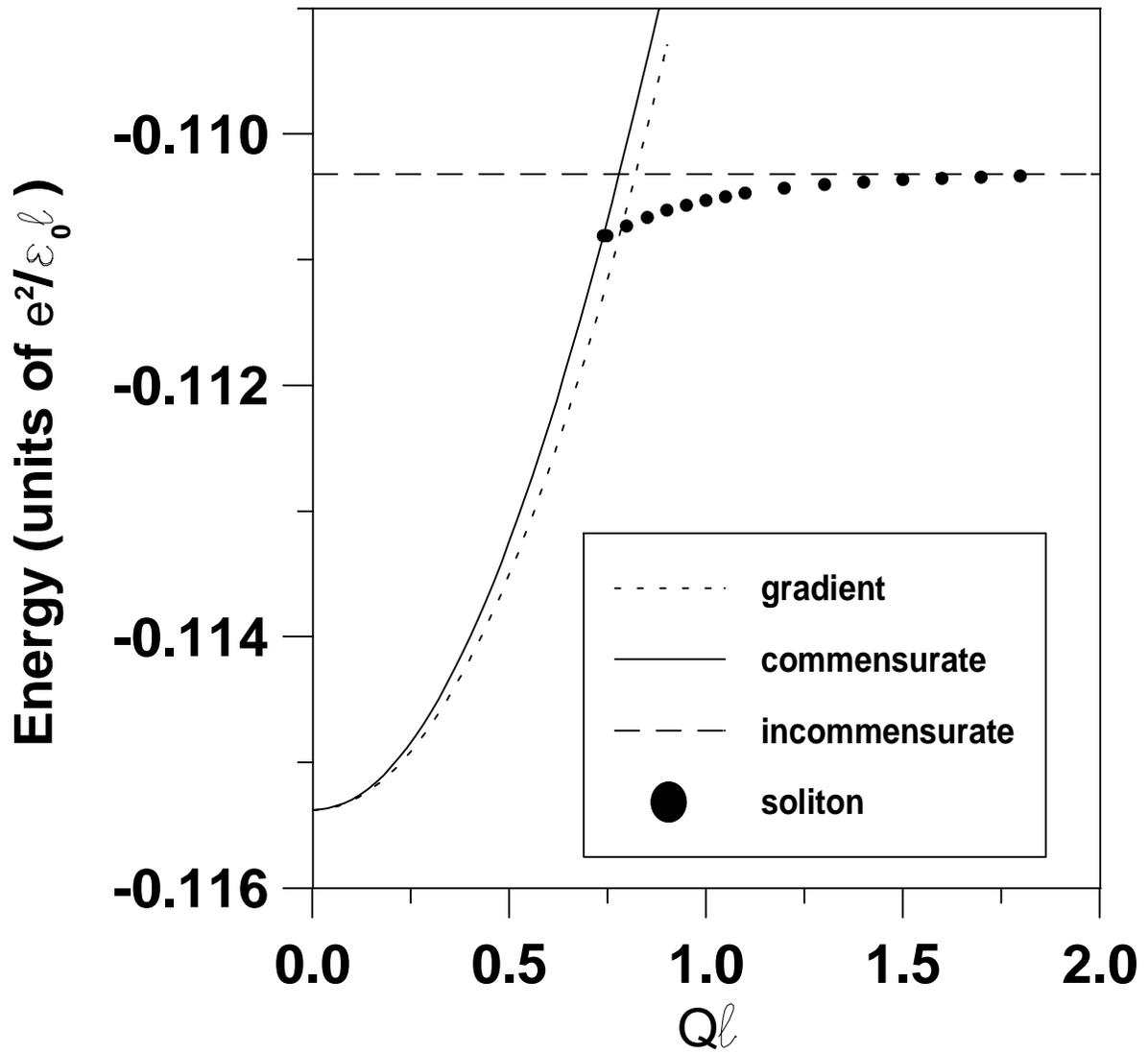
\begin{figure}
\caption{Energy of the commensurate, incommensurate and soliton-lattice
states as a function of $Q\ell $ for (a) $d/\ell =1.0,t/\left(
e^2/\epsilon_0\ell
\right) =0.01$ and (b) $d/\ell =1.877,t/\left(e^2/\epsilon_0\ell\right)
=0.005$. The
dotted line corresponds to the energy of the commensurate phase calculated in
the
gradient approximation.}
\end{figure}

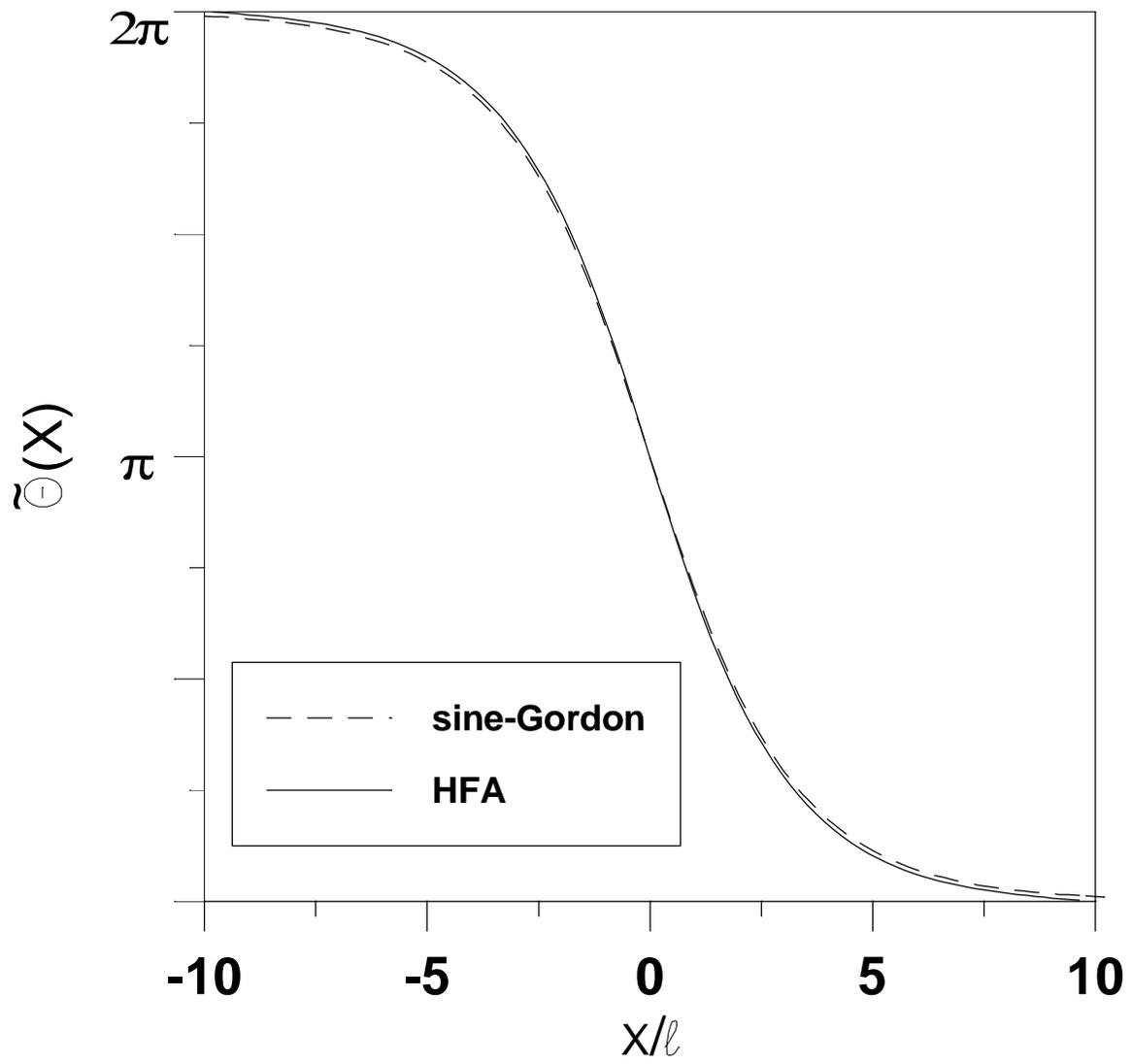
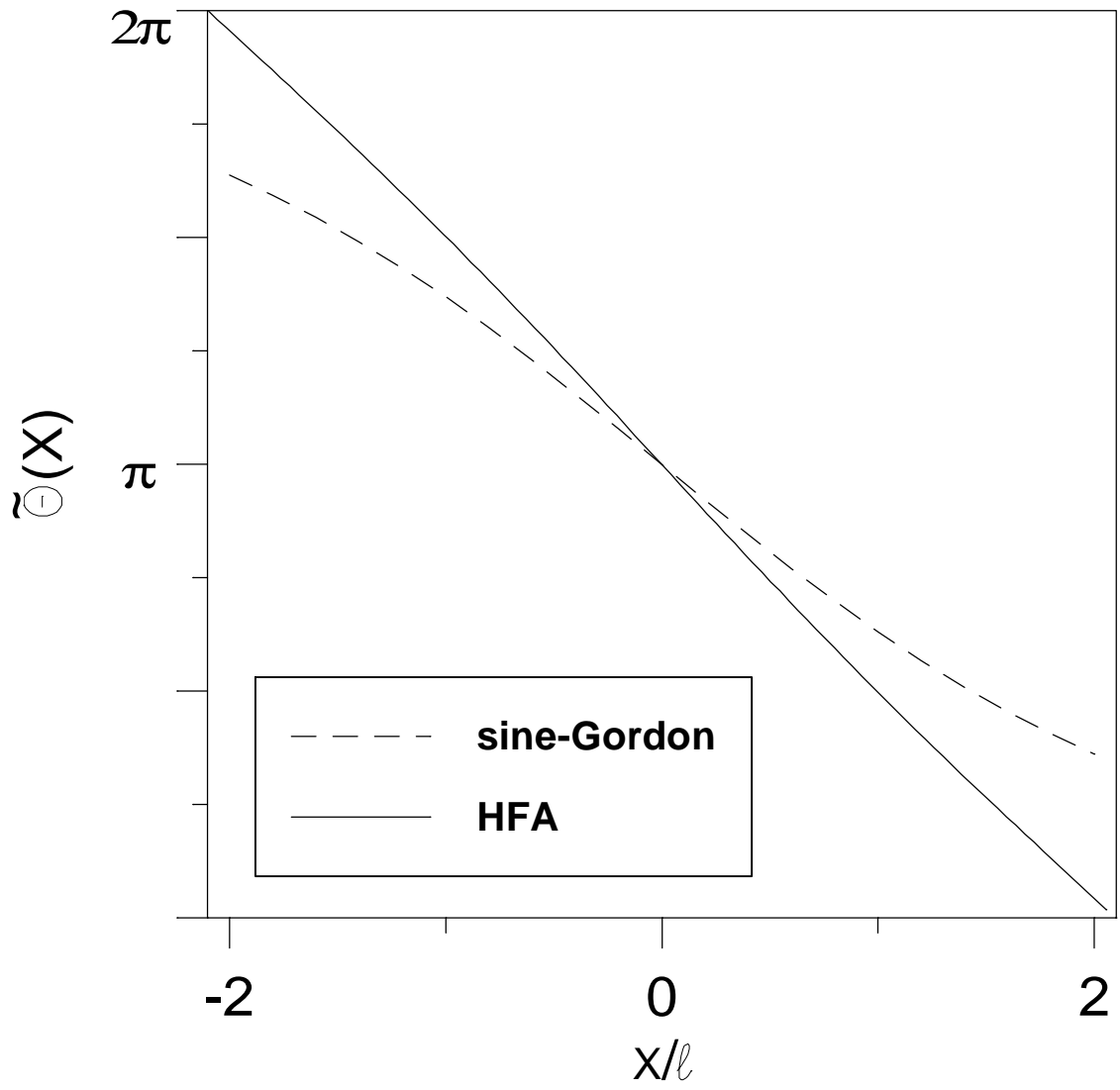
\begin{figure}
\caption{Behavior of $\widetilde{\theta }(X)$ over one period of
the soliton lattice in the HFA.
(a) $d/\ell =1.0,\ t/\left( e^2/\epsilon_0\ell \right) =0.01, Q\ell =0.63$
(just after the
$C\rightarrow S$ transition); (b) $d/\ell =1.0, t/\left( e^2/\epsilon_0\ell
\right)
=0.01, Q\ell =1.5$ (in the high
 parallel magnetic field limit). The dotted
 lines represent the sine-Gordon solution for
$\widetilde{\theta}(X)$
}
\end{figure}

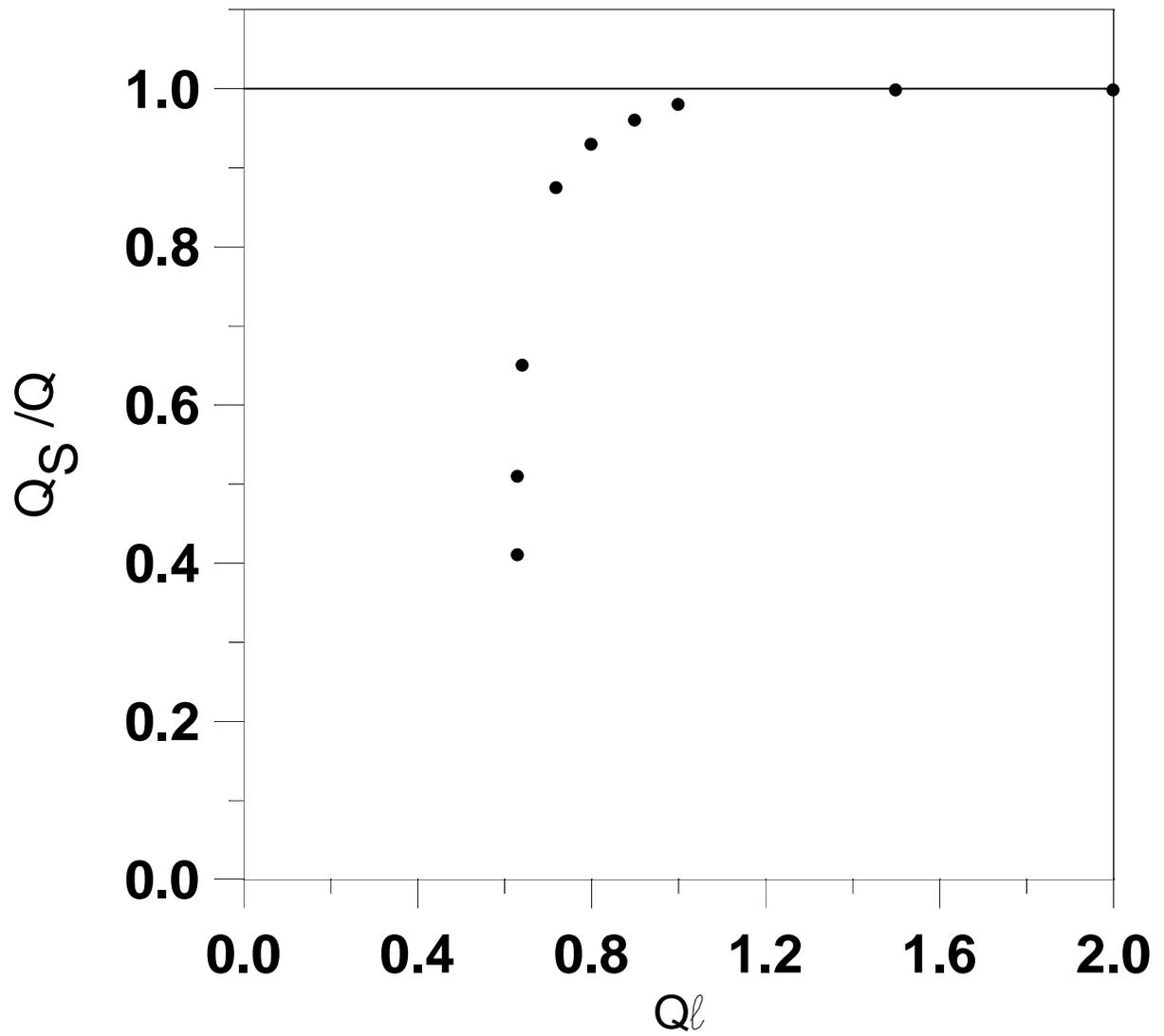
\begin{figure}
\caption{Behavior of the ratio
$Q_S/Q$ (soliton wave vector to parallel magnetic field wave vector)
 as a function of $Q\ell$ for $d/\ell =1.0,t/\left(
e^2/\epsilon_0\ell \right) =0.01$. The line at
$Q_S/Q=1$ is a guide to the eyes.
 }
\end{figure}

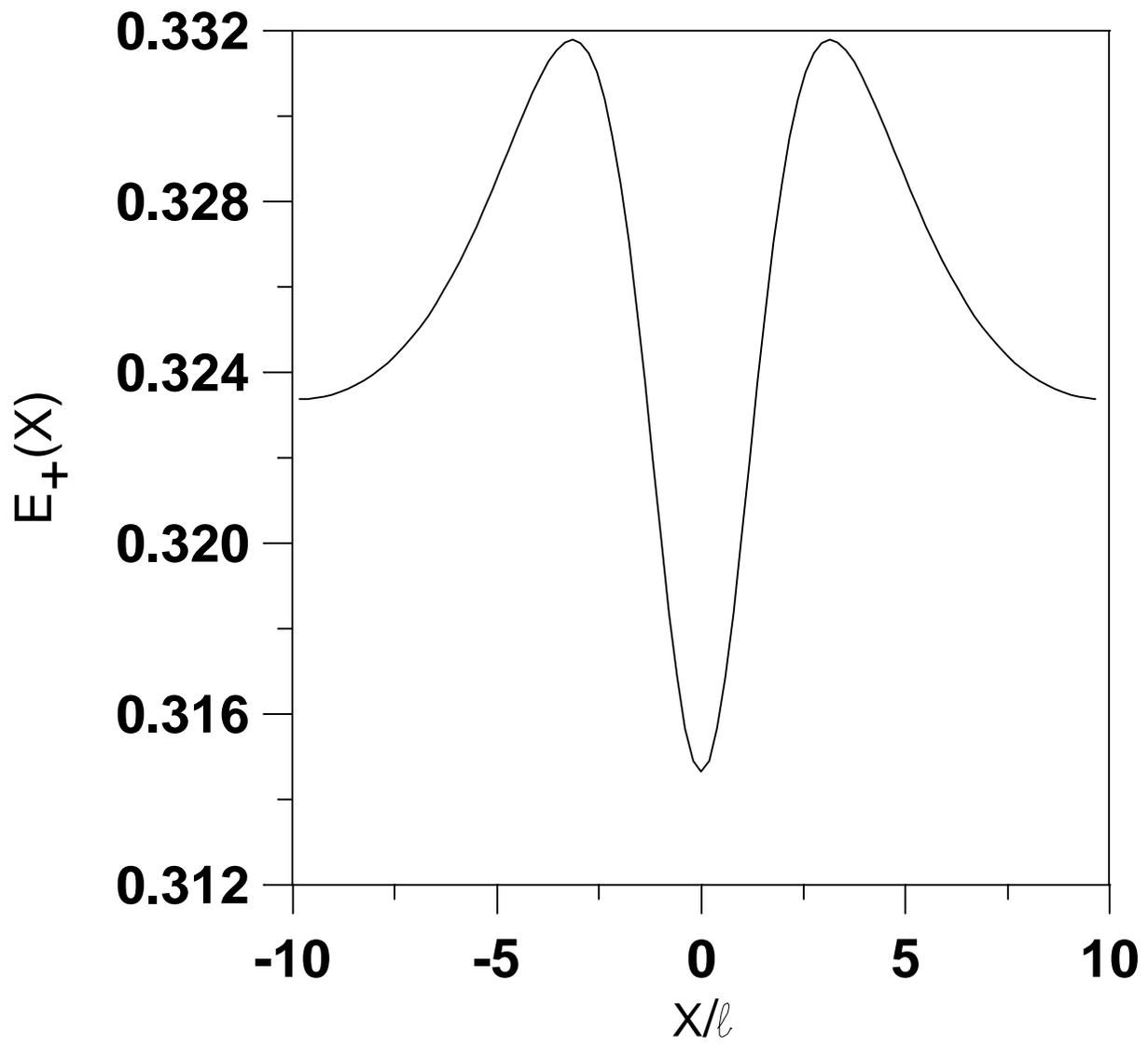
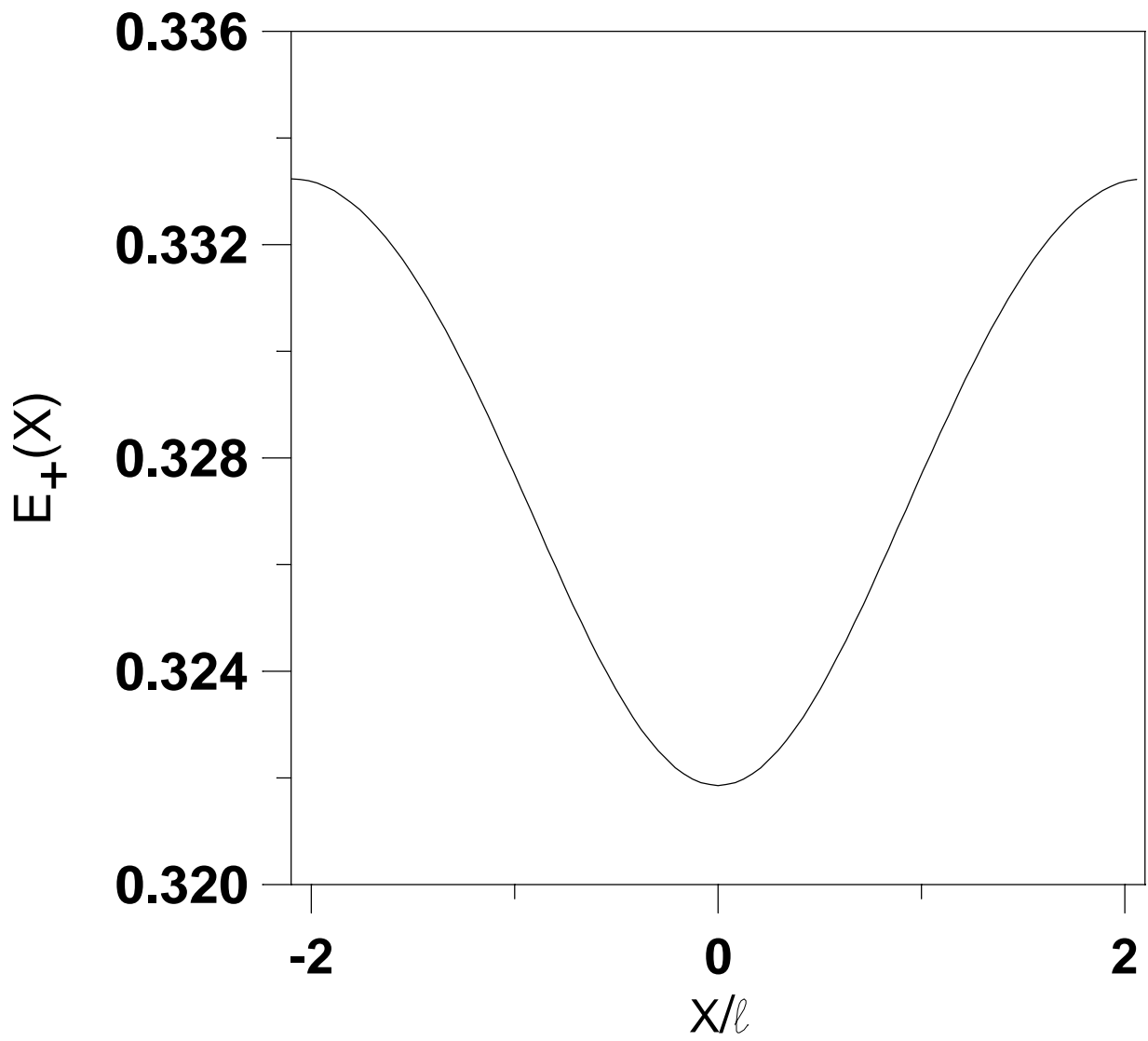
\begin{figure}
\caption{Shape of the upper band $E_{+}(X)$ over one period of the
soliton lattice in the
HFA for
(a) $d/\ell =1.0,t/\left( e^2/\epsilon_0\ell \right) =0.01, Q\ell =0.63$ and
(b) $%
d/\ell =1.0,t/\left( e^2/\epsilon_0\ell \right) =0.01, Q\ell =1.5$.}
\end{figure}

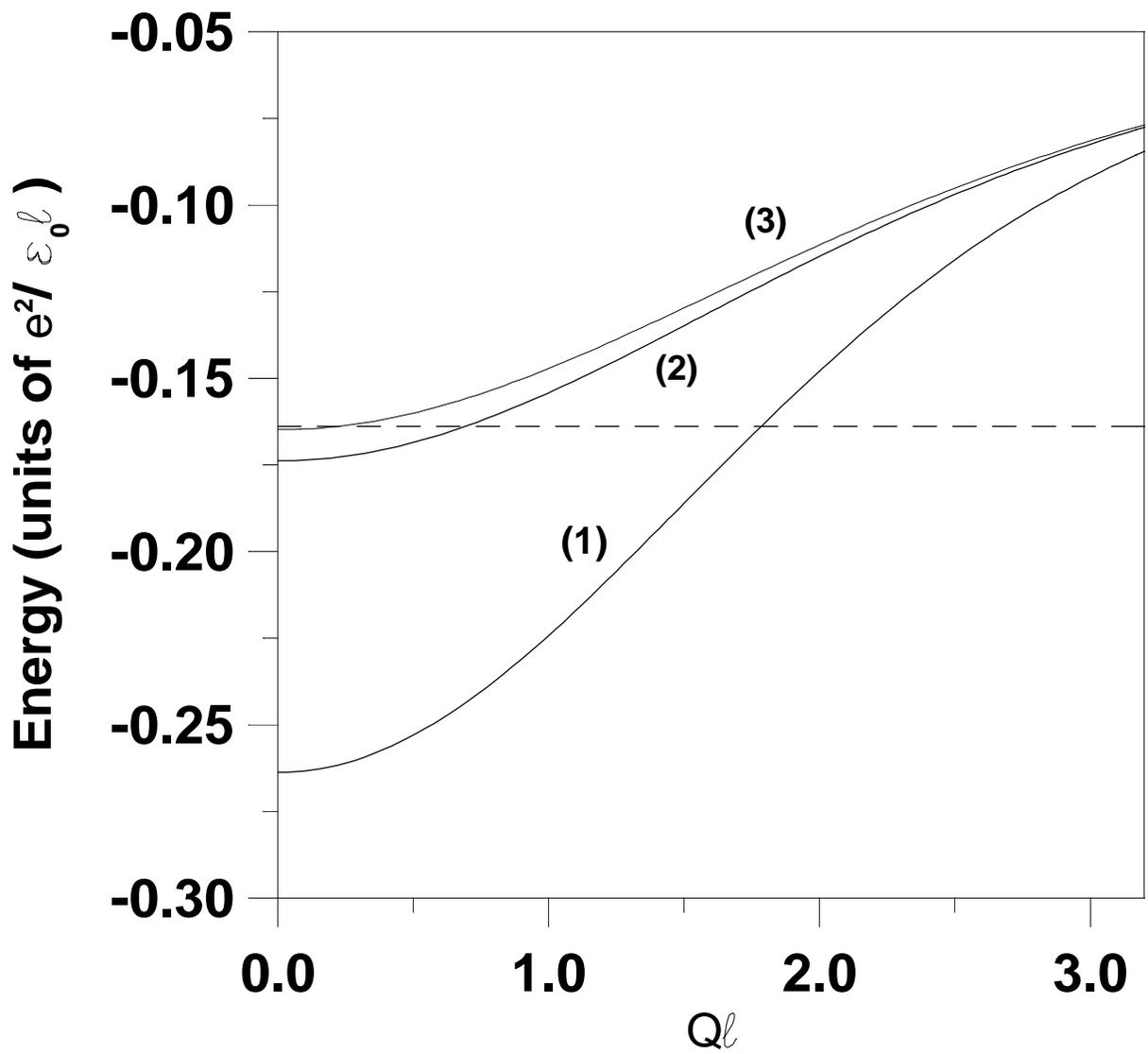
\begin{figure}
\caption{Energy of the commensurate state as a function of $Q\ell $
for different values of the bare hopping parameter $t.$
(1) $t/\left( e^2/\epsilon_0\ell \right) =0.1$;
(2) $t/\left( e^2/\epsilon_0\ell \right) =0.01$;
(3) $t/\left( e^2/\epsilon_0\ell \right) =0.001$.}
\end{figure}

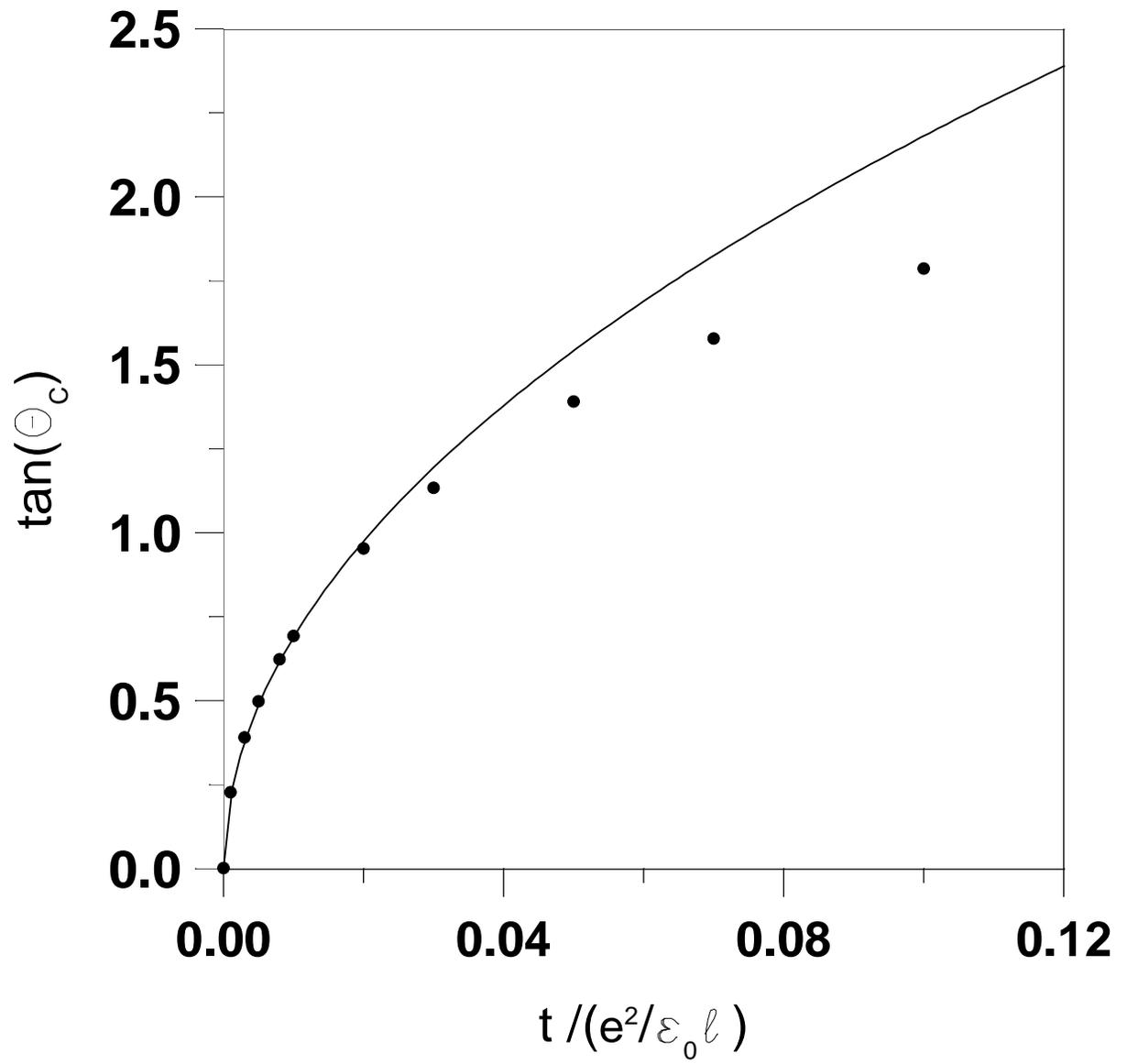
\begin{figure}
\caption{Critical tilt angle at which the transition from commensurate to
incommensurate states appears as a function of the bare hopping for $d/\ell
=1.0.$ The full line is a fit to a square root dependence.}
\end{figure}

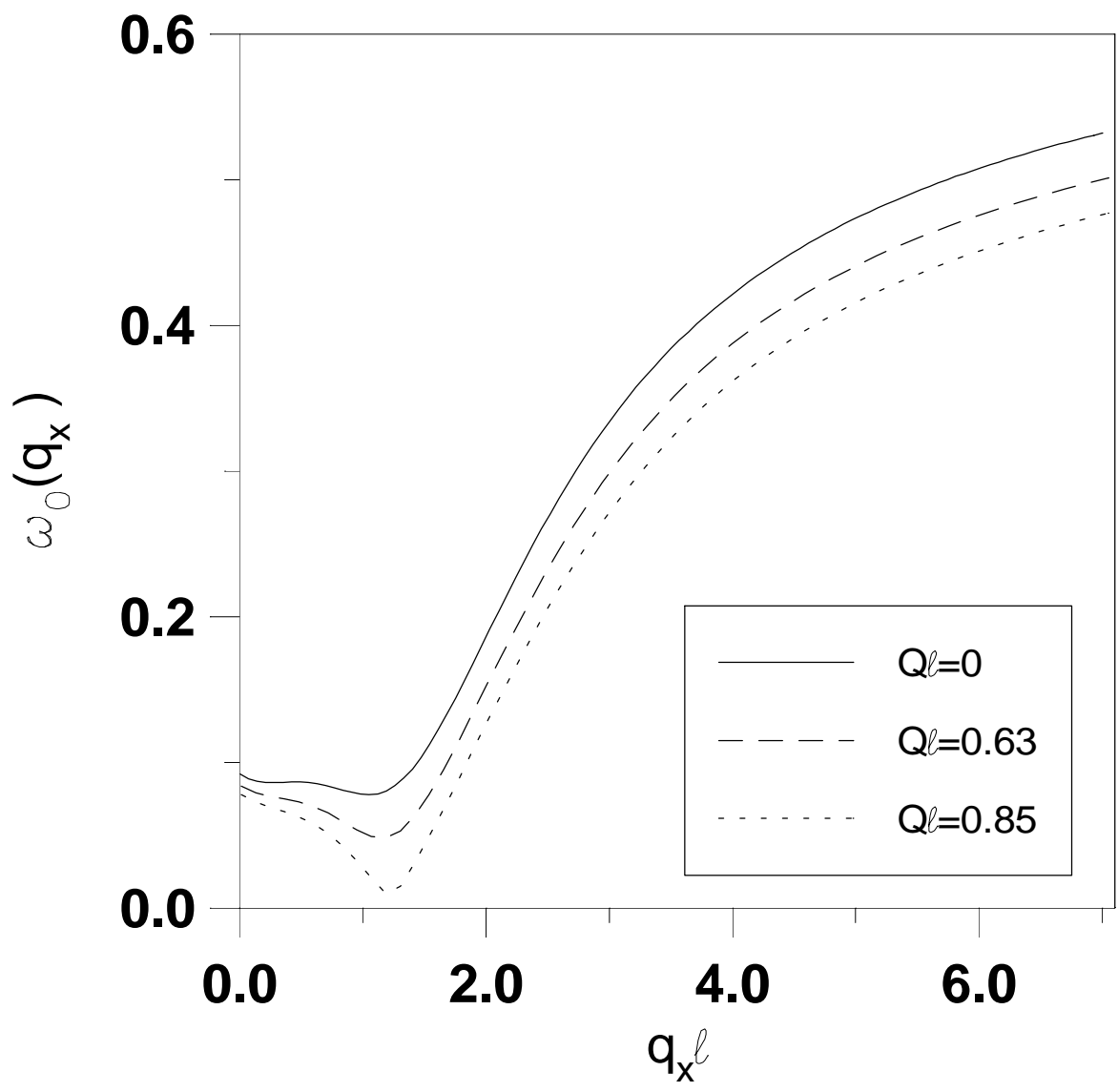
\begin{figure}
\caption{Dispersion relation of the spin wave mode in the
commensurate phase for $d/\ell =1.0,t/\left( e^2/\epsilon_0\ell \right) =0.01$
and
different values of the parallel magnetic field.}
\end{figure}

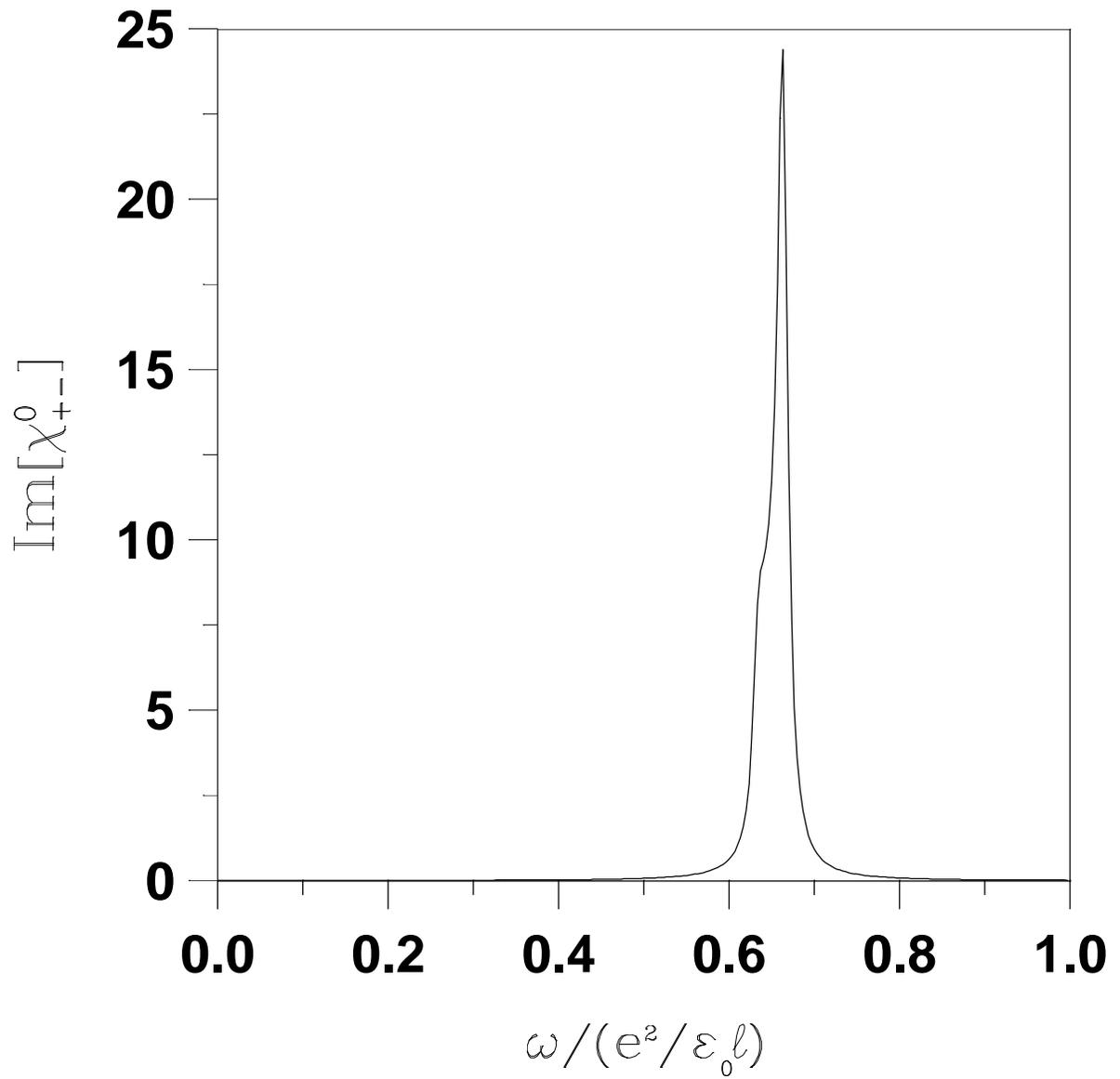
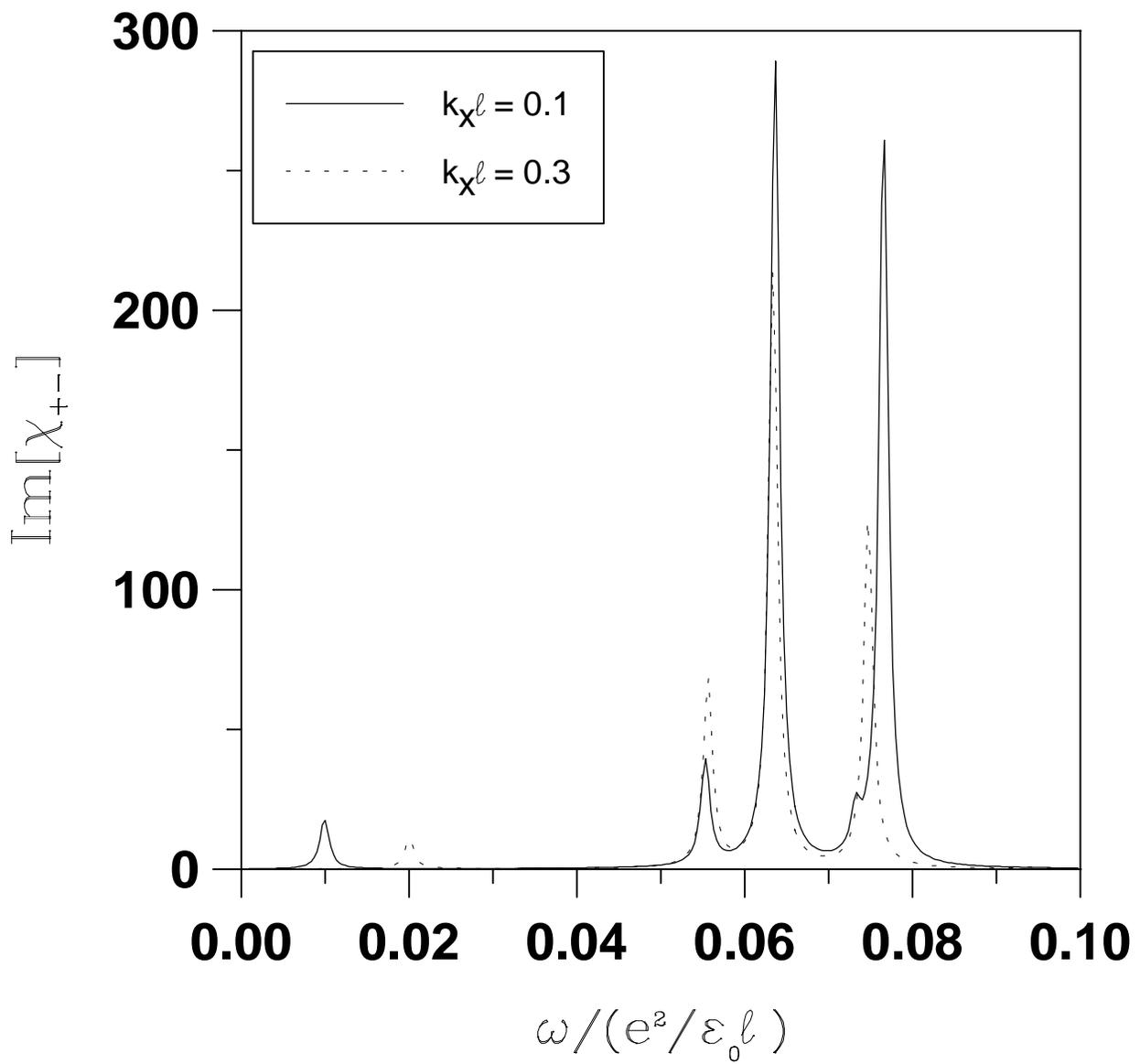
\begin{figure}
\caption{Imaginary part of $\chi_{+-}({\bf k}+{\bf Q},{\bf k}+{\bf Q},\omega)$
for $d/\ell =1.0$,$t/\left( e^2/\epsilon_0\ell\right)
=0.01$, $Q\ell =0.72$ and wave vectors
${\bf k}\ell=(0.1,0),(0.3,0)$. (a) HFA and (b) GRPA.
The Goldstone modes appear as low-energy peaks in
(b).}
\end{figure}

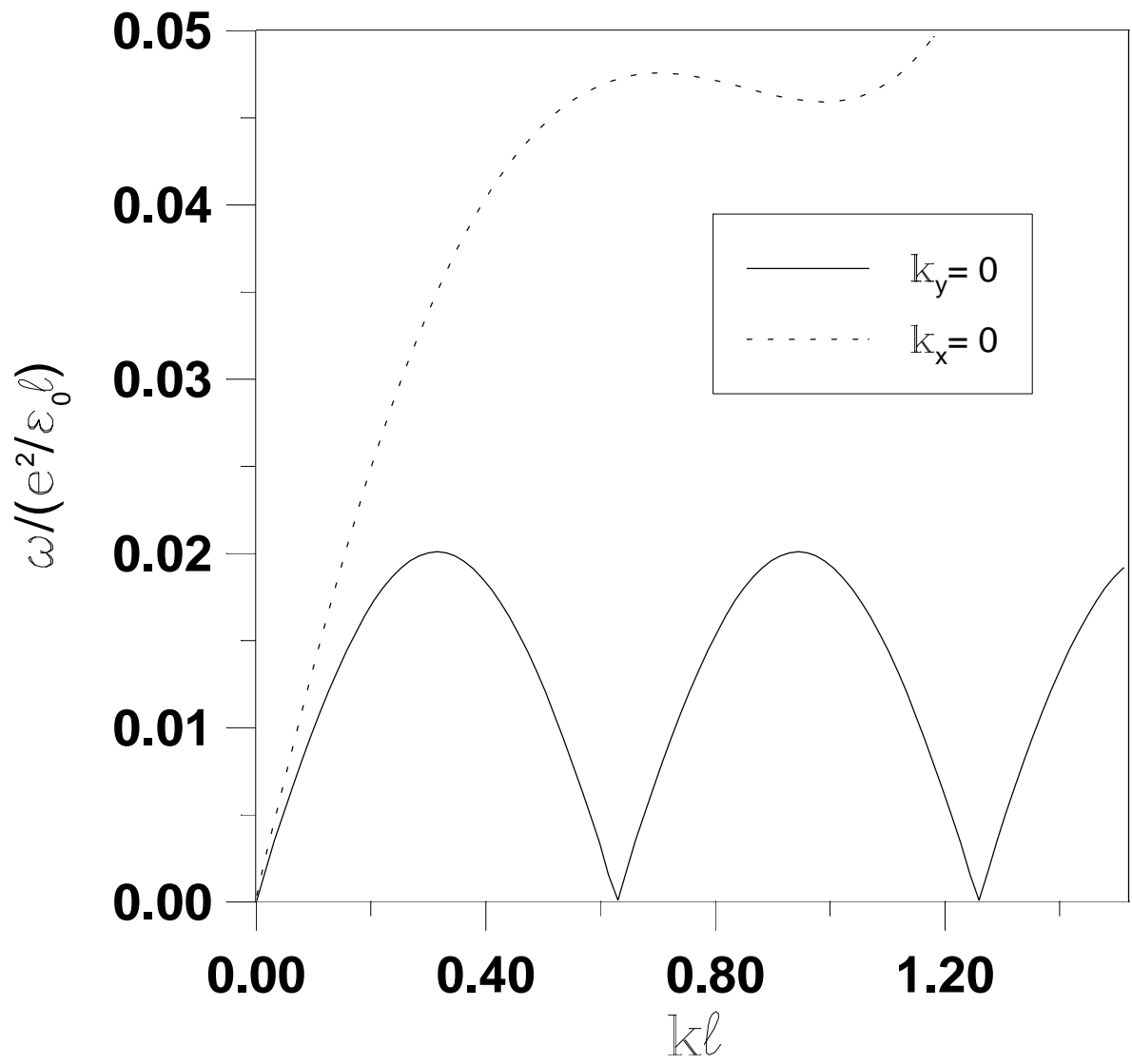
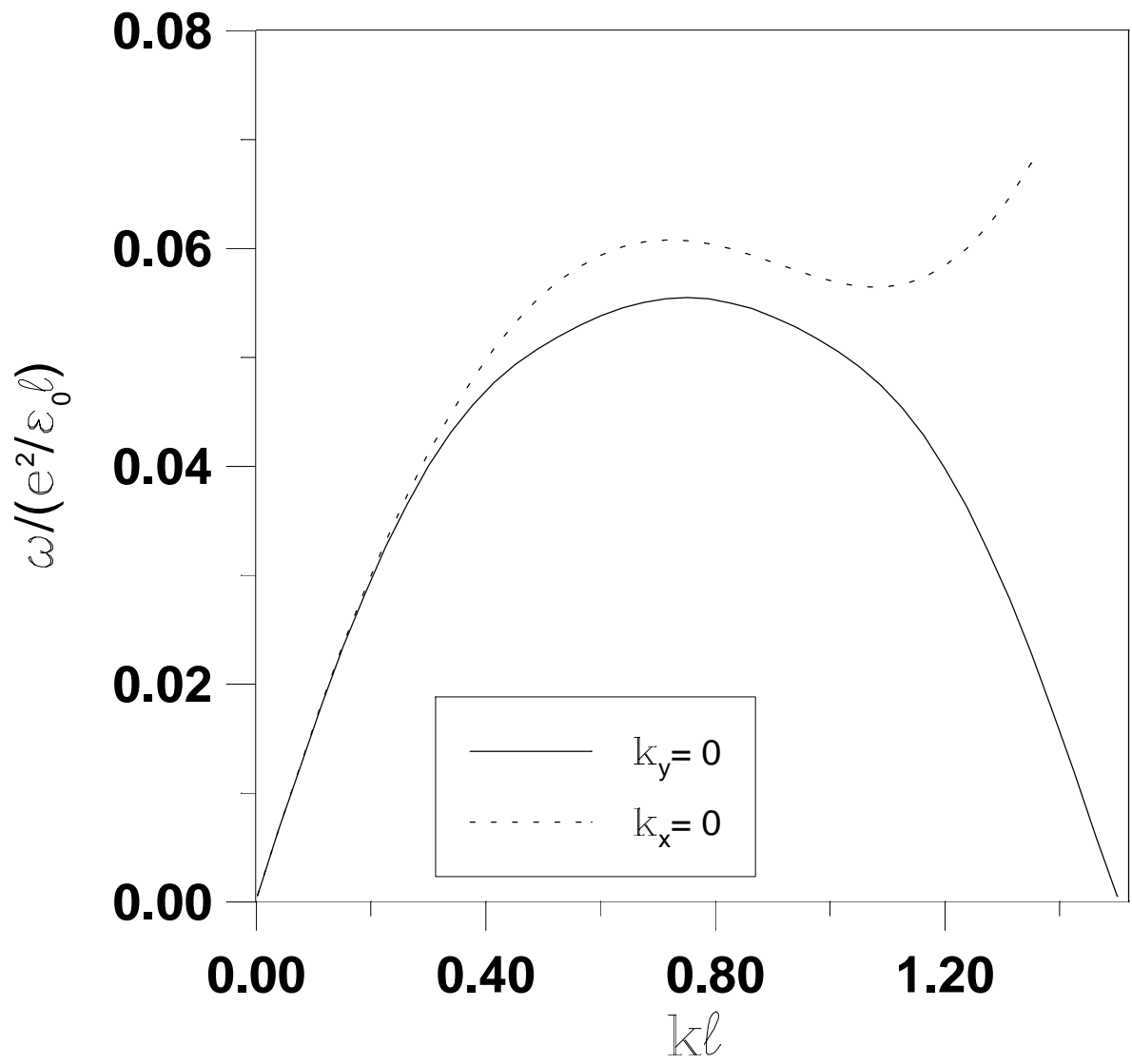
\begin{figure}
\caption{Dispersion relations of the Goldstone mode at (a) $d/\ell
=1.0$,$t/\left( e^2/\epsilon_0\ell \right) =0.01$, $Q\ell =0.72,$
and (b) $d/\ell
=1.0$, $t/\left( e^2/\epsilon_0\ell \right) =0.01$,
 $Q\ell =1.5$ along the directions $k_x=0$
and $k_y=0$. The dispersion relation is highly anisotropic, being in fact
periodic only along $k_y=0.$}
\end{figure}

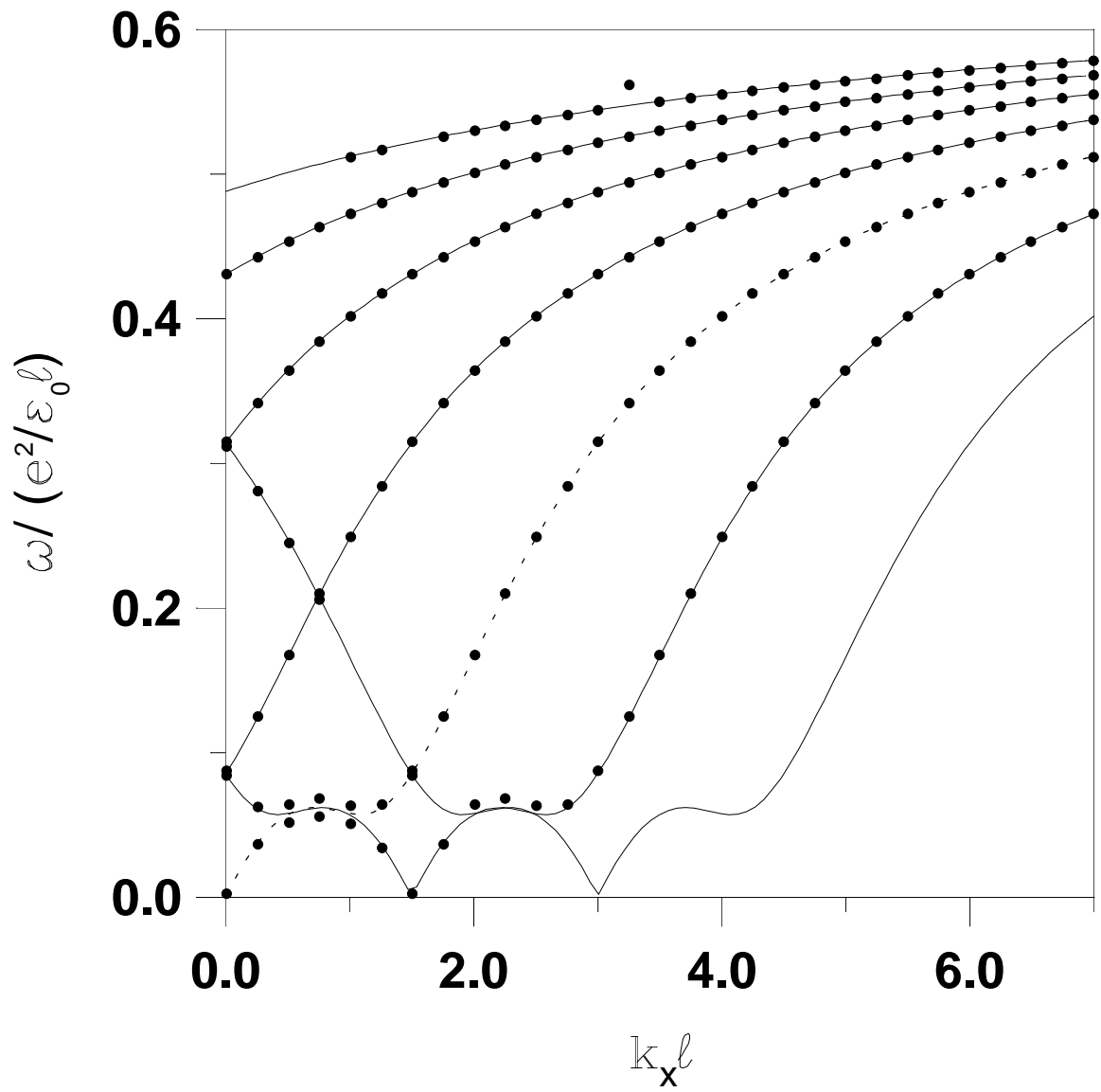
\begin{figure}
\caption{Comparison between the collective modes of the soliton lattice
($\bullet$)
at, $t/\left( e^2/\epsilon_0\ell \right) =0.01$,
in the high-parallel magnetic field regime and the collective modes
of the spontaneously-coherent state in the absence of tunneling (lines). The
parameters are
$d/\ell =1.0$, $Q\ell=1.5$ and
$k_y=0$.
For
the soliton lattice, only the strongest six modes
in the response function $\chi_{+-}$ are shown. For the spontaneously-coherent
state, $\omega_{sc}({k_x+nQ})$ is
plotted for $n=-2,-1,...,4$. The dotted line represents
$\omega_{sc}({k_x} )$. }
\end{figure}

\vfill\eject

\appendix
\section*{The gradient approximation}

In this appendix, we derive the Hartree-Fock approximation
equation of motion for the
single-particle Green's functions $G_{i,j}(X,\tau )=-<Tc_{i,X}(\tau
)c_{j,X}^{\dagger }(0)>$ instead of for its Fourier transform as in Eq.(\ref
{aonze}). We then introduce the so-called gradient approximation developed
in Ref.\cite{kunyang}. This approximation is very helpful in understanding how
the system evolves from the commensurate to incommensurate phases in the
presence of a parallel magnetic field.

In $X$ space, the Hartree-Fock Hamiltonian of Eq.(\ref{aneuf}) is
\begin{equation}
\label{bun}
\begin{array}{rcl}
H & = & -
\widetilde{t}\sum_X\left( e^{-iQX}\rho _{RL}(X)+e^{iQX}\rho _{LR}(X)\right)
\\  & - & \frac 1g\left( \frac{e^2}{\epsilon _0\ell }\right)
\sum_{X,X^{\prime }}K(X-X^{\prime })\left[ <\rho _{RL}(X)>\rho
_{LR}(X^{\prime })+<\rho _{LR}(X)>\rho _{RL}(X^{\prime })\right] ,
\end{array}
\end{equation}
where we again assume that the DQWS is unpolarized and neglect the constant
first term in Eq.(\ref{aneuf}).  We have also defined the operator
\begin{equation}
\rho _{i,j}(X)=c_{i,X}^{\dagger }c_{j,X},
\end{equation}
and the function
\begin{equation}
K(X-X^{\prime })=\sum_{q_x}V_d(q_x)\cos \ \left[ q_x(X-X^{\prime })\right] .
\end{equation}
The equations of motion for the single-particle Green's functions $%
G_{i,j}(X,\tau )$ are
\begin{equation}
\label{bdeux}
\begin{array}{rcl}
(i\omega _n+\mu )G_{RR}(X,\omega _n)+S^{*}(X)G_{LR}(X,\omega _n) & = & 1, \\
(i\omega _n+\mu )G_{LR}(X,\omega _n)+S(X)G_{RR}(X,\omega _n) & = & 0,
\end{array}
\end{equation}
where
\begin{equation}
J(X)=\frac 1g\left( \frac{e^2}{\epsilon _0\ell }\right)
\sum_{X^{\prime }}K(X-X^{\prime })<\rho _{RL}(X^{\prime })>,
\end{equation}
and
\begin{equation}
S(X)=\widetilde{t}e^{iQX}+J(X).
\end{equation}
Uncoupling Eqs.(\ref{bdeux}), we have
\begin{equation}
G_{RR}(X,\omega _n)=\frac{\frac 12}{i\omega _n+\mu -\left| S\left( X\right)
\right| }+\frac{\frac 12}{i\omega _n+\mu +\left| S\left( X\right) \right| },
\end{equation}
and
\begin{equation}
G_{LR}(X,\omega _n)=\frac 12\left( \frac{S(X)}{\left| S\left( X\right)
\right| }\right) \left[ \frac{-1}{i\omega _n+\mu -\left| S\left( X\right)
\right| }+\frac 1{i\omega _n+\mu +\left| S\left( X\right) \right| }\right] .
\end{equation}
There are two bands with energies $\pm \left| S\left( X\right) \right| $. At
$T=0K$, one of these bands must be entirely filled with the other one empty in
order to insure that the filling factor of the DQWS be exactly one. Thus,
\begin{equation}
<\rho _{RL}(X)>=\frac 12\left( \frac{S(X)}{\left| S\left( X\right) \right| }%
\right) \equiv \frac 12e^{i\theta (X)},
\end{equation}
or,

\begin{equation}
\label{btrois}\tan \theta (X)=\frac{\Im \left[ S(X)\right] }{\Re \left[
S(X)\right] }.
\end{equation}
This is the Hartree-Fock equation for the order parameter in $X$ space. We
have devised an iterative scheme to solve this equation and at the same time
minimize the Hartree-Fock ground-state energy per particle which is given by

\begin{equation}
\label{bquatre}E=-\frac{\widetilde{t}}g\sum_X\cos \left[ \theta
(X)-QX\right]
 -\frac 1{4g^2}({\frac{e^2}{\epsilon_0\ell}})
\sum_{X,X^{\prime }}K(X-X^{\prime })\cos \
(\theta (X)-\theta (X^{\prime })).
\end{equation}
We obtain results which are identical to those obtained using
the approach described in the body of this paper.

It is easy to see that Eq.(\ref{btrois}) can be obtained by minimizing the
energy with respect to the phase {\it i.e.} with $\frac{\delta E}{\delta
\theta \left( X\right) }=0$. The Hartree-Fock solution is thus an extremum
of the Hartree-Fock energy as expected.

The function $K(X-X^{\prime })$ is a rapidly decreasing function of
$|X-X^\prime|$.
If the angle $\theta (X)$ does not vary
too rapidly in space, this suggests that we can expand the cosine function in
Eq.(\ref{bquatre}). We get in this way
\begin{equation}
E\simeq -\frac 14\left( \frac{e^2}\ell \right) V_d(0)-\frac{\widetilde{t}}g%
\sum_X\cos \left[ \theta (X)-QX\right] +\frac \gamma {8g^2}\sum_X\left(
\frac{d\theta (X)}{dX}\right) ^2,
\end{equation}
where $\gamma =\left( \frac{e^2}{\epsilon _0\ell }\right) \sum_XK(X)X^2.$
Going to the continuum limit with $\sum_X\ldots \longrightarrow \frac{L_y}{%
2\pi \ell ^2}\int dX\ \ldots $, we have finally for the {\bf total} energy
of the system:
\begin{equation}
\label{bcinq}E_T\simeq -\frac g4\left( \frac{e^2}\ell \right) V_d(0)+L_y\int
dX\left[ \frac 12\rho _S\left( \frac{d\widetilde{\theta }(X)}{dX}+Q\right)
^2-\frac{\widetilde{t}}{2\pi \ell ^2}\cos (\widetilde{\theta }(X))\right] ,
\end{equation}
where $\widetilde{\theta }(X)=\theta (X)-QX$ and
\begin{equation}
\label{ros}
\begin{array}{rcl}
\rho _S & = & \left(
\frac{e^2}{\epsilon_0\ell} \right) \frac{L_y}{16g\pi ^2\ell ^4}\int dX\
K(X)X^2, \\  & =
& -\left( \frac{e^2}{\epsilon_0\ell} \right) \frac 1{8\pi \ell ^2}\left.
\frac{d^2V_d(Q)%
}{dQ^2}\right| _{Q=0}.
\end{array}
\end{equation}

If we now minimizes this total energy with respect to $\widetilde{\theta }%
(X),$ we find the well-known sine-Gordon equation
\begin{equation}
\label{bsix}\frac{d^2\widetilde{\theta }(X)}{dX^2}=\frac{\widetilde{t}}{2\pi
\rho _S\ell ^2}\sin (\widetilde{\theta }(X)),
\end{equation}
which admits the kink soliton as a solution:
\begin{equation}
\label{bsept}\widetilde{\theta }(X)=4\tan ^{-1}\left[ e^{-\sqrt{\frac{%
\widetilde{t}}{2\pi \rho _S\ell ^2}}X}\right] .
\end{equation}
If we insert Eq.(\ref{bsept}) in Eq.(\ref{bcinq}), we find that the
energy per particle in the presence of a single soliton is
\begin{equation}
E_{SS} \simeq E_C+L_y\rho _S\left[ 8\sqrt{\frac{\widetilde{t}}{2\pi
\rho _S\ell ^2}}-2\pi Q\right] .
\end{equation}
The first term in the right-hand-side of this equation is the
energy per particle
in the commensurate phase given in Eq.(\ref{nou}) while the
second term is the change in this energy due to the presence of the soliton.
When this second term becomes less than zero, we expect a phase transition
in the system to occur since then the energy to create one soliton becomes
negative. This happens at a parallel magnetic field given by
\begin{equation}
\label{bhuit}B_{||_{(C\rightarrow S})}=B_{\perp }\left( \frac{2l}{^{\pi d}}%
\right) \sqrt{\frac{2\widetilde{t}}{\pi \rho _S}}.
\end{equation}
If we define the width of the soliton as
\begin{equation}
\label{xsi}\xi =\frac 1{\sqrt{\frac{\widetilde{t}}{2\pi \rho _S\ell ^2}}},
\end{equation}
then, just at the transition, we have
\begin{equation}
\xi _c=\frac 4{\pi Q}.
\end{equation}
The critical magnetic field given in Eq.(\ref{bhuit}) should be compared
with the critical field for the transition from the commensurate to the
incommensurate phase. In the gradient approximation, this critical field is
given by given by
$$
B_{||(C\rightarrow I)}=\left( \frac \pi {2\sqrt{2}}\right) B_{\perp }\left(
\frac{2\ell }{\pi d}\right) \sqrt{\frac{2\widetilde{t}}{\pi \rho _S}.}
$$
Finally, for a periodic ground state with regularly spaced
solitons, the number of solitons is given by $N_S=L_x/L_S$ where $L_S$
is the distance between two solitons. If we neglect the soliton-soliton
interaction, then the energy of the configuration is approximately given
by
\begin{equation}
E\simeq E_C+\rho _S(Q_S\ell)
\left[ 8 \xi \ell -2\pi Q\ell\right] ,
\end{equation}
where $Q_S=2\pi/L_S$.  For $ Q > 4 /( \pi \xi) $ this energy is always
lowered by increasing $Q_S$ and hence the density of solitons.
Repulsive interactions between the solitons must be included to
determine the equilibrium value of $Q_S$.


\begin{references}
\bibitem{comment1} In the general case we can define a Fourier transform
which depends on two wavevectors.  See Refs. \cite{cotemacdo} and
\cite{cotebrey}.
\end{references}
\end{document}